\documentclass[final,5p,times,twocolumn]{elsarticle}

\journal{XXX}

\usepackage{lineno}
\modulolinenumbers[5]

\usepackage{hyperref}
\makeatletter
\g@addto@macro{\UrlBreaks}{\UrlOrds}
\makeatother

\usepackage[dvipsnames]{xcolor}

\usepackage[shortcuts]{extdash}

\usepackage{subcaption}

\usepackage{float}
\usepackage{epstopdf}
\usepackage{graphics}
\DeclareGraphicsExtensions{.eps,.gz,.eps,.pdf,.png,.jpg}
\usepackage{graphicx}

\usepackage{paralist}
\usepackage{dsfont}

\usepackage{amsthm}
\usepackage{amsmath}
\usepackage{amssymb}

\usepackage{multirow}
\usepackage{booktabs} 
\usepackage{mathtools}
\makeatletter \newif\if@restonecol \makeatother  
\usepackage[linesnumbered, ruled, vlined]{algorithm2e}
\usepackage{algpseudocode}

\newcommand{\sstitle}[1]{\smallskip\noindent\textbf{#1.\/}}

\newcommand{\sititle}[1]{\smallskip{\it #1:\/}}

\DeclareMathOperator*{\argmax}{arg\,max}

\newtheorem{problem}{Problem}
\newtheorem{theorem}{Theorem}

\theoremstyle{Definition}

\def\Snospace~{\S{}}

\makeatletter
\newcommand{\removelatexerror}{\let\@latex@error\@gobble}
\makeatother

\begin{document}

\begin{frontmatter}

\title{Model-Agnostic and Diverse Explanations for Streaming Rumour Graphs}

\author[griffith]{Thanh Tam Nguyen}
\author[griffith]{Thanh Cong Phan}
\author[hust]{Minh Hieu Nguyen}
\author[humboldt]{Matthias Weidlich}
\author[uq]{Hongzhi Yin}
\author[griffith]{Jun Jo}
\author[griffith]{Quoc Viet Hung Nguyen}

\address[griffith]{Griffith University, Australia}
\address[hust]{Hanoi University of Science and Technology, Vietnam}
\address[humboldt]{Humboldt-Universit\"at zu Berlin, Germany}
\address[uq]{The University of Queensland, Australia}

\begin{abstract}
The propagation of rumours on social media poses an important threat to 
societies, so that various techniques for rumour detection have been proposed 
recently. 
Yet, existing work focuses on \emph{what} entities constitute a rumour, 
but provides little support to understand \emph{why} the entities have been 
classified as such. This prevents an effective evaluation of the detected 
rumours as well as the design of countermeasures.
In this work, we argue that explanations for detected rumours may be given in 
terms of examples of related rumours detected in the past. A diverse set of  
similar rumours helps users to generalize, i.e., 
to understand the properties that govern the detection of rumours. Since the 
spread of rumours in social media is commonly modelled using feature-annotated 
graphs, we propose a query-by-example approach that, given a rumour graph, 
extracts the $k$ most similar and diverse subgraphs from past rumours. 
The challenge is that all of the computations require fast assessment of similarities between graphs.
To achieve an efficient and adaptive realization of the 
approach in a streaming setting, we present a novel graph representation 
learning technique and report on implementation considerations.
Our evaluation experiments show that our approach 
outperforms baseline techniques in delivering meaningful explanations 
for various rumour propagation behaviours. 

\end{abstract}


\end{frontmatter}


\section{Introduction}

\begin{figure}[!h]
  \centering
  \includegraphics[width=0.65\linewidth]{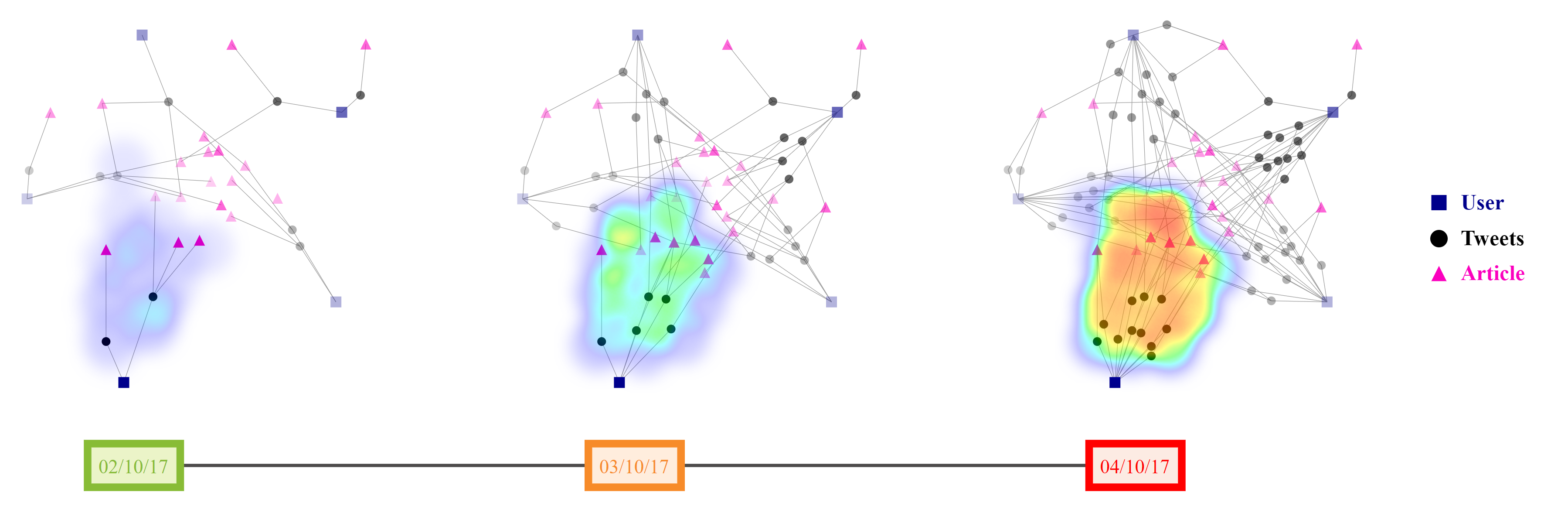}
  \caption{Exemplary propagation structure of a rumour.}
  \label{fig:example}
\end{figure}

Rumours on social media are an increasingly critical issue, threatening  
societies at a global scale. They fuel misinformation campaigns, reduce the 
credibility of news outlets, and potentially affect the opinions and, hence, 
actions of large parts of a society. The mechanisms for content sharing and 
user engagement that form the basis of most social media platforms foster the 
propagation, amplification, and reinforcement of rumours in graph-based 
propagation structures~\cite{vosoughi2018spread,tam2019anomaly,9119847}.
\autoref{fig:example} illustrates such a propagation structure for a rumour 
that emerged about the Las Vegas shooting in 2017, which attracted millions of 
engagements on Twitter and news websites.

With rumours emerging as a worldwide phenomenon, 
various techniques for their detection have been 
proposed~\cite{prasojo2019news,tam2019anomaly,lakshmanan2019combating}. 
However, existing techniques focus on identifying \emph{what} entities are 
part of a rumour structure, but provide little support for explaining the 
obtained results~\cite{reis2019explainable}. Especially the increasing 
popularity of methods for rumour detection based on deep learning, which 
exploit latent features, calls for explanations on \emph{why} certain entities 
have been classified as a rumour. 
Such explanations are important for various reasons:
(1) they enable an evaluation of the detected rumours and, hence, serve to 
assess the accuracy of rumour detection techniques; (2) they outline 
opportunities for improving the models adopted by the detection techniques; and 
(3) they enable users to devise and implement countermeasures that slow down 
and prevent the spread of rumours in the future.

Explanations of rumours may be based on different paradigms.
Since rumours are shared by humans, it is difficult to extract 
discrete rules to serve as explanations, though~\cite{yang2019xfake}. 
Feature-based 
explanations, in turn, cannot capture the graph-based propagation structures of 
rumours~\cite{shu2019defend}. Also, counter-factual approaches are not suited 
to understand why entities are classified as being part of a 
rumour, due to their focus on features that may change the 
result~\cite{hiep3404835cf}.
As such, explanations shall be based on a set of related examples, which 
enable users to generalize their properties~\cite{mathov2022not}. For a 
rumour represented as a graph, the explanation is then given in terms of 
related subgraphs of rumours detected in the past.

However, existing example-based explanation methods are still limited by easily overlooked assumptions. First, it is often assumed that the underlying model is stable over time. In practice, ML models are routinely retrained to react to changes in the overall environment and to changes in users' features. Explanations are often generated by a model-specific approach to identify the decision boundary of a specific model~\cite{mothilal2020explaining}; and thus, become obsolete quickly when the model changes. Second, it is often assumed that there is no constraint on model access and complexity. Most of existing works generate synthetic data samples and calling the model evaluation on these samples repeatedly to identify the decision boundary to find most similar ones to the original input. However, the model is not always available for such evaluation calls (e.g. software and hardware requirements) and can be very costly in practice (e.g. some models might take minutes, if not hours, to complete a prediction)~\cite{barocas2020hidden}.

To tackle these issues, we follow a model-agnostic approach to generate examples that are applicable to different types of models. In particular our approach works generally well with various problems and domains and is free from model access requirements. The performance of existing model-specific approaches are not generalized to other settings and even change considerably with a small of change of experimental setups~\cite{barocas2020hidden}. Moreover, our approach is not limited to a single example as in many previous works~\cite{hiep3404835cf}, but include diverse examples to increase the likelihood of returning at least one option that matches the user's true interest. Especially, our approach is free of model bias, which can still happen in many modern ML models~\cite{wang2022debiased,mazumder2022protected}.

A realization of example-based explanations for rumour detection 
faces diverse challenges. First, social media platforms are highly dynamic. 
The characteristics 
of rumours evolve over time, hence explanations are instance-specific. 
Second, rumours on social media materialize along 
multiple modalities, making traditional textual analysis 
insufficient~\cite{yang2019xfake}. For example, a rumour may be characterized 
by the relations between users and their interactions, not only by some textual 
content. 
Third, 
the representation of rumours as graphs induces computational challenges. It is  intractable to query the respective graph structures 
directly~\cite{duong2021efficient}.

In this paper, we address the above challenges and derive explanations for 
rumours with a query-by-example approach. Given a rumour graph, 
it extracts the $k$ subgraphs of past rumours that are most useful to explain 
the rumour, thereby achieving instance-specific explanations. By adopting a 
model based on 
feature-an\-no\-ta\-ted graphs, diverse modalities can be incorporated. 
Moreover, we 
present a novel graph representation 
learning technique, which enables us to derive explanations efficiently in a 
streaming setting. More specifically, our contributions can be summarised as 
follows:
\begin{compactitem}
	\item We define the problem of example-based explanation for rumour 
	detection. This is the first formulation of explanations for dynamic, 
	graph-based rumours.
	\item We provide a framework to address the explanation problem. 
	Given a rumour, we show how to score the utility of 
	subgraphs of past rumours and select the most useful ones.
	\item We design a graph representation learning technique to enable an 
	efficient computation of explanations.
	\item We outline how the efficiency and robustness can be 
	increased by indexing and caching schemes.
	\item We propose an adaptation mechanism to concept drifts, i.e. when new rumour patterns emerge and are different from past rumour examples.
\end{compactitem}
We further report on evaluation experiments with synthetic and
real-world data. We show that our approach delivers meaningful 
explanations for various rumour propagation behaviours. Compared to baseline 
techniques, selection of explanations with our approach is $4\times$ faster, 
while the runtime for graph similarity scoring improves by up to four orders of 
magnitude. Our approach also accumulates more than $16\times$ higher utility of 
explanations when a large number of rumours are detected and streamed into the 
system.

In the remainder, \autoref{sec:related} summarizes related work, before 
\autoref{sec:model} introduces a graph-based model for our work. 
The problem of example-based explanations for rumour detection and our general 
approach for it are introduced in \autoref{sec:approach}. 
We first clarify the utility scoring and selection of subgraphs in 
\autoref{sec:explanation}, before realizing these steps based on graph 
embeddings in \autoref{sec:rumour_embedding} and presenting further 
optimizations in \autoref{sec:extension}. We report on experiment results in 
\autoref{sec:exp}, before we conclude in \autoref{sec:conclusion}.

\section{Related Work}
\label{sec:related}

\noindent
\textbf{Explainable rumour detection.}
Most existing approaches to rumour detection neglect the question \emph{why} 
certain entities have been classified as a rumour. For this reason, techniques 
from the field of explainable machine learning have been adopted recently. 
However, explainable rumour detection focuses on pinpointing specific features 
that explain the detection results~\cite{reis2019explainable}. Also, for rumour 
detection based on deep learning, attention mechanisms and gradient activation 
maps have been employed to explain the latent features used in the 
detection~\cite{kumar2020fake}. However, understanding the structure of rumours 
turned out to be more important than understanding the detection 
itself~\cite{zhou2020survey,huang2019deep,huang2020heterogeneous,huang2020deep}. Due to the graph-based propagation structures of 
rumours, explanations solely based on features insufficient~\cite{nguyen2017retaining,hung2017computing,nguyen2020entity,nguyen2021structural}.

Recently, approaches to construct graph-based explanations using graph neural 
networks have been proposed~\cite{ying2019gnnexplainer,9119847,song2022bi,song2021jkt,xue2022dynamic}. However, the explanation is 
still given in terms of gradient footprints of the deep learning model rather 
than the characteristics of the rumour itself. Also, the 
explanation technique depends on the deep learning model and, hence, 
cannot be generalised to other detection techniques.  

We go beyond the state of the art by considering the detection technique as a 
black-box and by providing subgraphs of related rumours as explanations. We 
later show empirically that such explanations have a high discriminating power.

\sstitle{Computational explanations}
Various paradigms to explain computational results have been studied in the 
literature, including interpretable models, rule-based explanations, 
counterfactuals, and 
feature-based explanations~\cite{zhang2022explainable}. Interpretable models enable a direct inspection of 
the factors that govern some computation, e.g., regressions, decision trees, 
generalized additive models, and Bayesian networks. For example, the weights 
learned in a linear regression and the path taken in a decision tree can 
directly be interpreted by a user. 
However, such interpretable models 
are not sufficiently expressive to capture the structure of rumours.

As argued above, explanations based on examples are most suited in the context 
of rumour detection. The reason being that rumours 
involve human perception, which is difficult to capture with discrete 
rules~\cite{yang2019xfake}, whereas feature-based approaches fail to capture 
the graph-based structures of rumour propagation~\cite{shu2019defend}. 
Counter-factual explanations, in turn, focus on features that may change the 
result of rumour detection rather than the features that led to the detection 
result in the first place~\cite{hiep3404835cf}. While synthetic examples have the advantage of gradient optimisation~\cite{yizhao2022interspeech} to generate them, real examples have the benefits of acting as case studies to convince decision makers in critical applications such as healthcare and social informatics~\cite{phan2022exrumourlens}.

Similar to example-based explanations based on subgraphs, as targetted in our 
work, exemplary contextual information, such as 
user comments, may be extracted to explain detected 
rumours~\cite{shu2019defend}. However, 
such contextual information tends to be subjective and provides limited insight 
on the rumour as a whole. By extracting subgraphs of past rumours, our approach 
supports a more comprehensive assessment of the properties that characterize a 
rumour. 

\sstitle{Subgraph querying}
Our work is orthogonal to graph search. This is because the subgraphs that 
serve as explanations do not necessarily participate in the prediction of the 
rumour of interest. Even if the subgraphs and the considered rumour are of 
similar structure (and they are close in the embedding space), it is not 
obvious to what extent the detection technique applied the same reasoning and 
which features influenced the classification as a 
rumour~\cite{tam2019anomaly}. Thereon, we formulate a 
subgraph selection problem that maximises a notion of explanation utility, 
which incorporates both similarity and diversity of the respective subgraphs.

\section{Model}
\label{sec:model}

Below, we introduce the formal model adopted in our work, which is also 
summarized in \autoref{tab:notations}. 

\sstitle{Multi-modal Social Graph (MSG)}
To capture social media data, we adopt the notion of a multi-modal social graph 
(MSG). It is an undirected graph $G=(V,E)$ with entities $V$ and relations $E 
\subseteq [V]^2$, where $[V]^2$ are all subsets of $V$ of size 
two~\cite{sun2011pathsim}. 

Each entity and relation has a type called modality. Formally, with 
$\mathcal{A}_v$ as a set of pre-defined entity modalities, we define a type 
function $\phi_{v}: V \rightarrow \mathcal{A}_{v}$ that maps each entity to 
a modality. An entity represents an instance of its 
modality. This model extends naturally to a streaming setting, where the set of 
entities $V$ is extended over time, but the set of modalities is fixed.
The modality of a relation $e=(v,v')$ is derived directly from the modalities 
of the entities, and captured by a type function $\phi_E: E 
\rightarrow [\mathcal{A}_v]^2$ with $[\mathcal{A}_v]^2$ being all subsets of 
$\mathcal{A}_v$ of size two. It is defined as $\phi_E(e) \mapsto  
(\phi_V(v),\phi_V(v'))$.
For brevity, we write $\phi(.)$ for either type function, when the 
domain is clear, and define $\mathcal{A} = \mathcal{A}_v \cup 
[\mathcal{A}_v]^2$ as the set of all entity and relation modalities.

Each entity or relation can be associated with a vector of features of its 
modality (aka intrinsic characteristics of the modality), such as language 
structures and lexical features of a social post~\cite{reis2019explainable}. 
These features often capture rumour signals in a social graph. 
In our model, we incorporate such features by a set of functions $f = 
\{f_1, \ldots, f_{|\mathcal{A}|}\}$, where $f_i: V \cup E \rightarrow 
\mathbb{R}^{d_i}$ assigns an $d_i$-dimensional feature vector $f_i(x)$ to each 
element $x \in V \cup E$ of modality $a_i \in \mathcal{A}$.

Social graphs can capture different types of social 
networks~\cite{fang2020effective}. In the remainder, we use the case of Twitter 
for illustration purposes. Then, the entities of an MSG are of the 
modalities user, tweet, hashtag, or link. Edges capture the relations between 
entities, e.g., that a user created a tweet or that a link was referred to in a 
tweet. 


%

\begin{table}[t]
	\centering
	\footnotesize
	\caption{Overview of important notations.}
	\label{tab:notations}
	\vspace{-0.5em}
	\begin{tabular}{ll}
		\toprule
		\textbf{Notation} & \textbf{Description} \\
		\midrule
		$\mathcal{A}=\{A_1, \ldots, A_m\}$       & A set of modalities     \\
		$\Omega = \cup_{A \in \mathcal{A}} A$ & All possible values of 
		modalities \\
		$G =(V,E)$ & A multi-modal social graph (MSG)\\
		$S = \{s_1, \ldots, s_k\}$ & An explanation of $k$ rumour subgraphs \\
		$\mu(S)$ & The utility of an explanation $S$ \\
		$\phi(v)$ & The modality of an entity $v$ \\
		$z_v$ &  Embedding of an entity $v$ \\
		$z_s$ & Embedding of a rumour subgraph $s$ \\
		\bottomrule
	\end{tabular}
	\vspace{-1.0em}
\end{table}

\sstitle{Rumours}
For the purpose of this paper, we largely abstract from the specifics of 
techniques for rumour detection (except in our evaluation experiments). That 
is, we assume that for a given multi-modal social graph, a rumour detector 
identifies rumours as subgraphs $\mathcal{S} = \{s_1, \ldots, 
s_n\}$ of the MSG. Moreover, in a streaming setting, the MSG is extended over 
time by the addition of entities and edges, so that also the rumour detector 
constructs sets of rumours over time.

\section{Problem Statement and Approach}
\label{sec:approach}

We first explain the intuition of example-based explanations 
under a graph-based model by means of an example in \autoref{sec:intuition}.
Then, we present a formal statement of example-based explanations in 
\autoref{sec:formulation} and outlined our general framework to address this 
problem in \autoref{sec:framework}.

\subsection{Intuition of Example-based Explanations}
\label{sec:intuition}

\begin{figure}[!h]
	\centering
	\vspace{-0.5em}
	\includegraphics[width=0.65\linewidth]{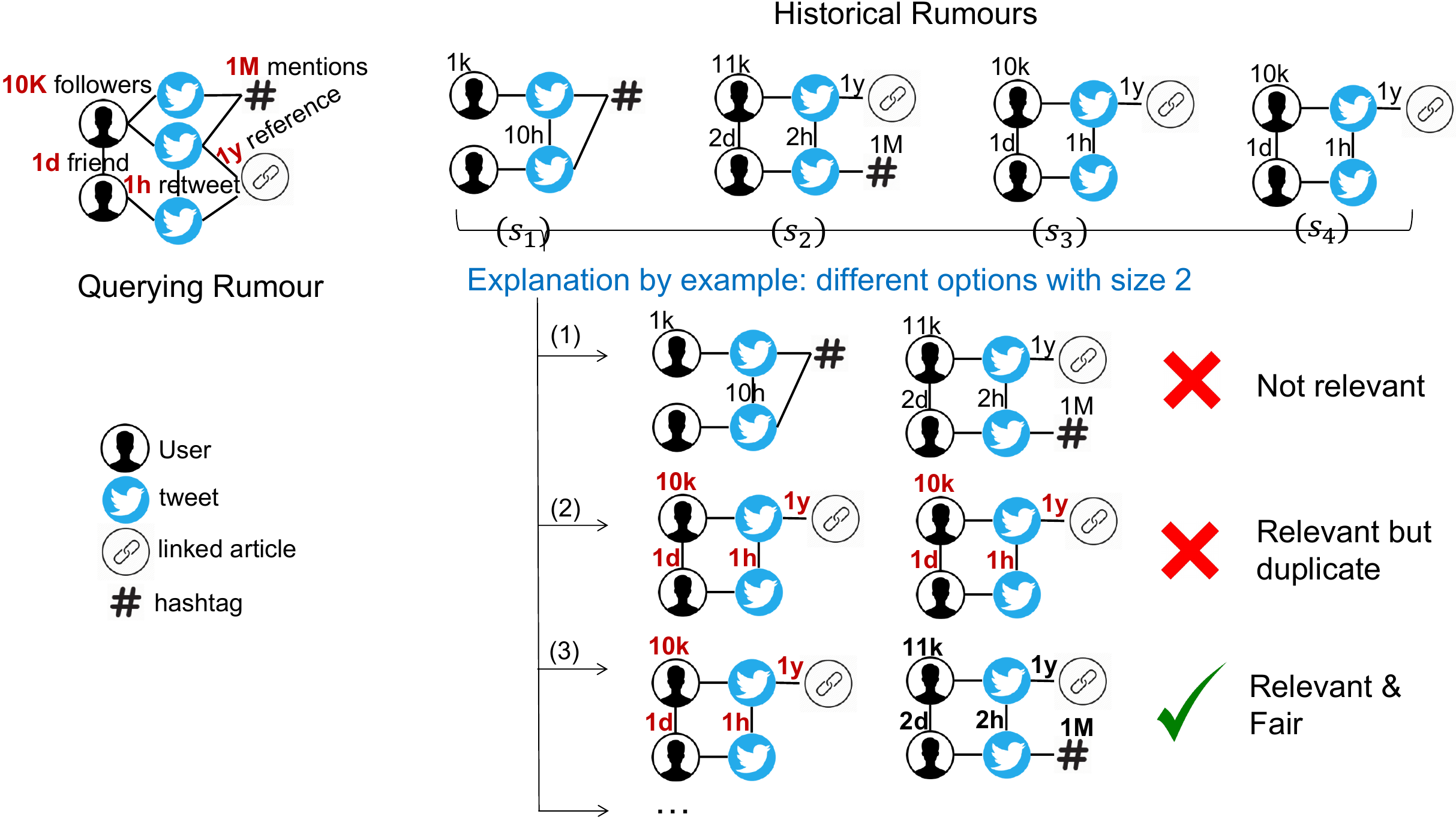}
	\vspace{-1em}
	\caption{Explanation by examples.}
	\label{fig:motivation}
	\vspace{-0.5em}
\end{figure}

Consider the example given in \autoref{fig:motivation}. Here, a given rumour 
(top left) 
includes several entities and relations, including specific users, tweets, a 
hashtag, and a linked article, each being assigned further features. For 
example, the first user has 10K followers and the relation between the two 
users is marked as a friendship link that has been established one day ago. 
The relation between two tweets indicates a retweet with a certain time 
difference (1 hour), while the hashtag has the number of 
mentions (1M) as a feature.

\autoref{fig:motivation} also shows previously detected 
rumours (top right), each differing in the structure and the 
associated features. Those that are similar to the rumour of interest may serve 
as explanations of rumour detection, as they enable users to generalize from 
the different structures and understand the general characteristics of the 
detected rumours. For 
example, in \autoref{fig:motivation}, users can deduce a pattern in the rumours 
that is given as a subgraph of a famous user (10k followers) with a new friend 
(1 day friendship relation), who posts a tweet that links to an old article 
(one year ago), which was quickly retweeted by the new friend.

Extracting a set of related rumours to explain a rumour of interest enables 
various downstream applications, such as \emph{group attack detection}, where 
the common subgraphs of the returned examples help to identify groups of 
bot and fake accounts~\cite{han2019efficient}; \emph{rumour diffusion}, where 
commonalities and differences of the examples characterize different 
propagation patterns~\cite{wang2015detecting}; and \emph{rumour 
source detection}, where invariant features of the examples help to identifying 
the genesis of rumours to develop mitigation and prevention 
mechanism~\cite{dong2019multiple}. 
To support these applications effectively, however, the extracted examples 
shall meet three criteria:
\begin{compactitem}
	\item \emph{Similarity:} Explanations shall be given through examples that 
	are similar to the rumour of interest. 
	For a graph-based model, similarity search is computational 
	challenging, though. 
	\item \emph{Diversity:} Explanations shall be fair in the sense that they 
	are not biased to one rumour structure. Information redundancy 
	shall be avoided to provide comprehensive insights~\cite{van2021evaluating}.
	\item \emph{Conciseness:} Explanations shall be as simple as possible, 
	following the principle of Occam's razor~\cite{yang2019xfake}. That is, 
	smaller sets of examples shall be favoured. 
\end{compactitem}

\subsection{Problem Formulation}
\label{sec:formulation}

The above criteria for explanations for rumour detection correspond to somewhat 
conflicting goals: similarity 
favours biased explanations; diversity favours different explanations; and 
conciseness favours a small number of examples and limits the amount of 
information. We model the trade-off between similarity and diversity and 
capture the desired information through a \emph{utility} function that assigns 
a score to all possible example (i.e., subgraphs of rumours in the past). 
Moreover, conciseness of the explanation is captured by a dedicated parameter 
that limits the number of extracted examples. Based thereon, we formulate the 
problem addressed in this work:  
\begin{problem}
Given 
a set of rumours detected in the past $\mathcal{S}$, a rumour of interest $q$, 
and a utility function $\mu: \mathcal{S} \rightarrow \mathbb{R}$, 
the problem of \emph{example-based explanations of rumour detection} is to 
extract a set $S^*\subseteq \mathcal{S}$ of $k$ rumours, such that:
\[
S^* = \argmax_{S \subseteq \mathcal{S}, |S| \leq k} \mu(S).
\]
\end{problem}
To support explainable rumour detection efficiently and effectively, a  
solution to extract the $k$ subgraphs of past rumours that are most useful 
shall further satisfy the following requirements:
\begin{compactitem}
	\item[(R1)] \emph{Efficient computation:} The utility of past rumours shall be 
	calculated fast with a small storage cost. 
	\item[(R2)] \emph{Incremental computation:} The approach shall be applicable for a 
	stream of detected rumours. 
	\item[(R3)] \emph{Adaptive computation:} The approach shall support efficient 
	updates of the underlying social graph.
\end{compactitem}

\subsection{Solution Framework}
\label{sec:framework}

\begin{figure}[!h]
	\centering
	\includegraphics[width=0.55\linewidth]{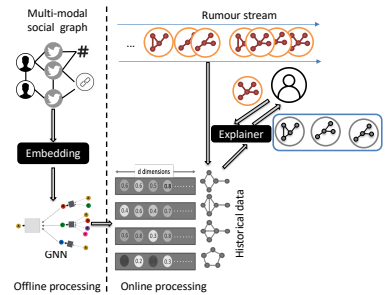}
	\caption{Framework for explainable rumour detection.}
	\label{fig:framework}
	\vspace{-1em}
\end{figure}

Our framework for example-based explanations of rumour detection is illustrated 
in \autoref{fig:framework}. As input, our framework takes a stream 
of detected rumours that are returned by some black-box detector. When not 
considering further optimisations, each detected rumour is stored as-is. 
When a user requests an explanation for a particular rumour, we select a set of 
$k$ subgraphs with maximal utility as an explanation (\emph{Explainer} in 
\autoref{fig:framework}). 
An instantiation of this approach requires the definition of the utility 
function to score the detected rumours and select the most useful ones. 
In \autoref{sec:explanation}, we provide such an instantiation of the utility 
function. Moreover, we propose several efficient selection algorithms, since 
the problem of selecting $k$ optimal subgraphs turns 
out to be NP-hard and a naive solution would require storing all rumour 
subgraphs. 

However, this first approach still has a high time complexity, as the 
similarity search requires expensive graph computations.
To achieve a more efficient extraction of explanations, we incorporate a novel 
technique for graph representation learning. First, we 
train an embedding model for the whole multi-modal social graph as part of an 
offline processing step (\emph{Embedding} in \autoref{fig:framework}). Then, 
when processing detected rumours in an online manner, each rumour is indexed as 
multi-dimensional vector by the learned embedding model. Once an explanation 
shall be derived for a rumour, we will, again, select a set of $k$ 
subgraphs with maximal utility (\emph{Explainer} in \autoref{fig:framework}). 
However, this optimal subset of rumours is now computed efficiently based on 
the subgraph's embeddings. 
As such, the challenge is to capture the characteristics of multi-modal social 
graphs so that the semantic similarity of rumours is reflected in 
the distance of their embeddings. 
The details of our technique for graph representation learning are given in 
\autoref{sec:rumour_embedding}.



\section{Utility Scoring \& Graph Selection}
\label{sec:explanation}

Below, we first discuss how to measure the similarity of multi-modal social 
graphs (\autoref{sec:similarity}) and the diversity of a set of graphs 
(\autoref{sec:diversity}). Subsequently, we incorporate both aspects in a 
utility score for explanations (\autoref{sec:goodness}). Finally, we address 
the selection of the $k$ most useful subgraphs 
(\autoref{sec:construct_explanation}).

\subsection{Quantifying Graph Similarity}
\label{sec:similarity}

Measures for graph similarity may follow two paradigms: exact 
subgraph matching (subgraph isomorphism) and inexact subgraph matching. In 
rumour detection, it is more appropriate to consider approaches of the latter 
category, since users are interested in slight variations of rumour 
structures. 
As such, our framework defines a retrieval problem, where the 
relevance of a subgraph $s\in \mathcal{S}$, given a subgraph $q$ for which an 
explanation shall be derived, is derived from the 
similarity of $s$ and $q$, denoted by $sim(s,q)$.
Various graph similarity measures for inexact graph matching have been proposed 
in the literature, see~\cite{yu2014subgraph}. 
In our framework, the 
graph similarity measure represents a design choice and the only assumption 
imposed is that the similarity score is normalized to $[0, 1]$. Specifically, 
we incorporate the following measures:
\begin{compactitem}
	\item \emph{MCS:} The size of the maximum common subgraph (MCS) between two 
	graphs is taken as a basic similarity measure. 
	This measure is used in many graph matching studies~\cite{zhu2013high}.
	\item \emph{Graphsim~\cite{yu2014subgraph}:} Graphsim generalizes the 
	concept of a MCS and incorporates data modalities. Based on an isomorphic 
	mapping between nodes of two 
	graphs, the measure compute the Jaccard similarity of their 
	modalities. The similarity of edges as well as the overall similarity of 
	two graphs are then derived from the similarity of the mapped nodes.
	\item \emph{GED:} The graph edit distance (GED) lifts the idea of the 
	string edit distance to graphs. It is often applied in error-tolerant graph 
	matching~\cite{duong2021efficient}. 
	\item \emph{Embedding-based similarity:} Graphs may be transformed to 
	multi-dimensional vectors using embedding~\cite{adhikari2018sub2vec} for similarity scoring. 
	That is, deep learning is leveraged to extract latent feature from both 
	structural and node/edge-level information. We later propose a 
	respective technique for multi-modal social graphs. 
\end{compactitem}
We illustrate graph similarity based on 
\autoref{fig:motivation} and the {MCS} measure, which counts the number of 
similar nodes and edges in the maximum 
common subgraph. We consider two entities or relations of 
the same modality to be similar, if they have the same features. For 
example, the user with 10K followers in the rumour of interest ($q$)  is 
similar to the user with 10K followers in rumour $s_3$. Then, $q$ and $s$ have 
five similar entities,  five similar relations, two different entities, and 
four different relations. Their graph similarity would be 
$sim(q,s_3)=\frac{5+5}{5+5+2+4} = 
0.625$. 
In the same vein, we have $sim(q,s_1) = 0.238$, $sim(q,s_2) = 0.273$, and 
$sim(q,s_4) = 0.625$.
%
In the remainder, we incorporate a similarity threshold $0 \leq \gamma \leq 1$ 
in the selection of subgraphs. Only subgraphs with a similarity score 
of at least $\gamma$ serve as candidates, i.e., $sim(s,q) \geq \gamma$, 
$\forall \ s \in 
S^*$. The threshold 
enables users to control the amount of structural differences between a rumour 
and the extracted examples.

\subsection{Quantifying Explanation Diversity}
\label{sec:diversity}

Next, we show how to achieve some diversity in the explanation 
constructed for a rumour. 
For graph data, a traditional definition of diversity, or fairness, relies on 
the coverage, the number of nodes of a graph that are part of a 
set of subgraphs. 
Such a definition 
is not suitable for multi-modal social graphs, though, 
where nodes may  
be virtually equivalent due to having the same neighbourhood structure, the 
same modality, and the same set of assigned features. Hence, considering a 
traditional notion of coverage for explanations would limit a user in 
understanding the characteristics of rumours.

\sstitle{Content-based coverage}
Our first idea to achieve a suitable notion of diversity is to eliminate 
explanations that contain many pairs of similar subgraphs. Given a set of $k$ 
subgraphs, $S=\{s_1, \ldots, s_k\}$, and a similarity measure $sim(.,.)$ for 
subgraphs, we define the content-based coverage of an explanation as:
\[
- \sum_{i=1}^k \sum_{j \neq i} sim(s_j,s_i).
 \]

\sstitle{Modality-based coverage}
Diversity of explanations shall further incorporate the modality of nodes. To 
this end, we first define the distance $dist(v,s)$ of a 
node $v$ to a subgraph $s=(V_s, E_s)$ as
\[
dist(v,s) = \min_{u \in V_s} dist(v,u)
\]
where $dist(v,u) = 1-sim(v,u)$ is the distance between nodes $u$ and $v$. 
We say that a modality $a \in \mathcal{A}$ is covered by a subgraph $s$, if 
there is a node $v$ with the same modality, i.e., $\phi(v) = a$, within distance $\delta$ from $s$, i.e., $dist(v,s) \leq 
\delta$, where $\delta$ is a distance threshold. 
By limiting the distance, we ensure that the modality plays a role in 
information propagation inside the subgraph~\cite{yu2014subgraph}.
Then, we 
define the influence of this modality as $\alpha^{dist(a,s)}$, where 
$dist(a,s)$ is the minimum distance of any node $v$ to $s$, with $\phi(v) = a$, 
and $\alpha$ being a decay factor. 
Based thereon, we define the modality-based coverage of a 
subgraph $s$ as:
\[
cov_\alpha(s) = \sum_{a \in \mathcal{A}: dist(a,s) \leq \delta} \alpha^{dist(a,s)}
\]
where $\alpha \in [0,1]$ and $\delta \geq 0$ are parameters that enable users 
to control this aspect of the coverage of the explanation. For example, 
increasing $\delta$ and $\alpha$ favours large subgraphs over smaller ones. 
 
 \subsection{Utility Scoring}
\label{sec:goodness}
 
Equipped with a similarity measure for subgraphs and notions of coverage to 
ensure diversity in a set of subgraphs, we are ready to define a measure for 
the overall quality of an explanation. 
Specifically, we propose three utility measures, which differ in the adopted 
notion of coverage. Given a rumour $q$ that shall be explained and a similarity 
measure $sim(.,.)$ for subgraphs along with a similarity threshold $\gamma$, 
the utility of a set of subgraphs $S\subseteq 
\mathcal{S}$ is assessed with one of the following measures:
\begin{compactitem}
	\item \emph{Content-based utility:}
\begin{equation}
	\label{eq:content_explainability}
	\mu_C(S) = 2 \sum_{s \in S} sim(s,q) - \lambda_1 \sum_{s}\sum_{s' \neq s} 
	sim (s,s')
\end{equation} 
where $\lambda_1 \leq \gamma / \max_{s \in S} \{\sum_{s' \in s} sim(s,s')\}$ is 
a balancing parameter between similarity and coverage. 
	\item \emph{Modality-based utility:}
\begin{equation}
	\label{eq:modality_explainability}
	\mu_M(S) = \sum_{s \in S} sim(s,q) + \lambda_2 \sum_{s \in S} sim(s,q) 
	cov_\alpha(s)
\end{equation}	
where $\alpha \leq \gamma$ and $\lambda_2$ is a balancing parameter again. 
Here, we add a factor $sim(s,q)$ to the coverage score to prevent selecting 
subgraphs of low relevance but with high diversity.
	\item \emph{Hybrid utility:}
\begin{equation}
\label{eq:hybrid_explainability}
\mu(S) = \mu_C(S) + \mu_M(S)
\end{equation}
\end{compactitem}
We illustrate the computation of $\mu_C$ with $\lambda_1=1$. Taking up the 
example from \autoref{sec:similarity}, we examine the utility of the second 
explanation, i.e., $S_2 = \{s_3,s_4\}$. We have $\mu_C(S_2) = 2*(0.625+0.625) - 
1 * 1 * 1 = 1.5$. Similarly, it holds $S_3=\{s_3,s_2\}$ and $\mu_C(S_3) = 
2*(0.625+0.273) - 1 * 0.467 * 0.467= 1.58$. Therefore, explanation $S_3$ would 
be preferred over $S_2$ in this configuration.

Before turning to the algorithms to select an explanation, we observe two 
important properties of $\mu_C$, $\mu_M$, and $\mu$.

\begin{theorem}[Monotonicity]
	Let $\mathcal{S}$ be a set of subgraphs of detected rumours. For any 
	subsets $S,S' \subseteq \mathcal{S}$ with $S \subseteq S'$, it holds that 
	$\mu_C(S) \leq \mu_C(S')$, $\mu_M(S) \leq \mu_M(S')$, and $\mu(S) \leq 
	\mu(S')$. 
\end{theorem}

%
%

\begin{theorem}[Submodularity]
	Let $\mathcal{S}$ be a set of subgraphs of detected rumours. For any 
	subsets $S,S' \subseteq \mathcal{S}$ with $S \subseteq S'$ and any subset 
	$R \subset \mathcal{S}$ with $R \cap S' = \emptyset$,  it holds that 
	$\mu_C(S \cup R) - \mu_C(S) \geq \mu_C(S' \cup R) - \mu_C(S')$. Similar 
	inequalities also hold for $\mu_M$ and $\mu$. 
\end{theorem}
Proofs of the properties of $\mu_C$ and $\mu_C$ are omitted due to space 
constraints. Monotonicity and submodularity of $\mu$ then follows from the 
preservation of these properties by 
summation~\cite{jin2021unconstrained,yu2014subgraph}.


%
%

\subsection{Selecting Optimal Explanations}
\label{sec:construct_explanation}

Given the above results and knowing that submodular maximisation 
is NP-hard~\cite{jin2021unconstrained}, we should not expect any 
exact, polynomial-time solution to construct an optimal explanation. Our goal 
therefore is to design algorithms to approximate a solution efficiently. Below, 
we first present a greedy 
selection strategy, before presenting a swap-based algorithm.

\sstitle{Greedy selection}
For submodular maximisation, there exists a standard greedy algorithm to select 
the top-$k$ subgraphs with an approximation ratio of $1-1/e 
\approx 0.632$~\cite{jin2021unconstrained}. It proceeds in two steps. First, we 
find all relevant subgraphs to 
$q$ with at least $\gamma$ similarity. Second, these 
subgraphs are ranked as follows. Let $S_k$ be the set of the top-$k$ subgraphs, 
which is initially empty. In each iteration of the greedy selection, a 
subgraph $s$ that maximises $\mu(S_k \cup \{s\}) - \mu(S_k)$ (or $\mu_C, 
\mu_M$, respectively) is selected. 

The time complexity of the greedy selection is $\mathcal{O}(k |\mathcal{A}| 
M)$, where $M$ is the total number of similar subgraphs, $|\mathcal{A}|$ is the 
number of  modalities, and $k$ is the size of selection. Since we need to store 
all similar subgraphs explicitly, the space complexity is $\mathcal{O}(M)$.

\sstitle{Swap-based selection}
A drawback of the greedy strategy is that we need to store all 
subgraphs and evaluate their relevance before selecting the top-$k$ results, 
which does not scale to large numbers of subgraphs. Moreover, each time 
a new rumour is detected, the selection needs to be recomputed from scratch.

Against this background, we devise a swap-based algorithm, inspired 
by Jin et al.~\cite{jin2021unconstrained}, that iterates over the candidate 
set multiple times. In each iteration, we maintain a set $S$ of $k$ subgraphs. We check if there exists a 
set $T$, where $T \cap S = \emptyset$ and $|T| \leq |S|$, and a set $S' 
\subseteq S$ such that $\mu(S \setminus S' \cup T) > \mu(S)$ (or $\mu_C, 
\mu_M$, respectively) and $|S'| = |T|$. 
If so, it will replace $S'$ by $T$.
While the algorithm runs in polynomial time, it has an approximate ratio of 
$1/2$. Supposing the algorithm goes 
through the entire set of subgraphs $P$ times, the time complexity becomes
$\mathcal{O}(k P |\mathcal{A}| M)$, which is worse than conducting 
the greedy selection $P$ times.
However, this mechanism hints us at the possibility of swapping a new 
subgraph with an old one without completely recomputing the result.

Extending the basic idea of swap-based selection, we evaluate each new subgraph 
on the fly, i.e., we only go through the entire set of subgraphs once without 
storing them. More precisely, we 
keep a set $C_k$ of $k$ subgraphs as a candidate solution. Each time a 
new rumour is detected, represented by subgraph $s$, we check if there exists a 
subgraph $s' \in C_k$ such that $\mu(C_k \setminus \{s'\} \cup \{s\}) > \beta 
\mu(C_k)$, for 
some $\beta \geq 1$. If so, we replace $s'$ with $s$. As such, $\beta$ denotes 
a trade-off parameter between the regret of removing old subgraphs 
and retaining new ones. 
Similar to \cite{jin2021unconstrained}, we can show that this one-pass swap-based 
selection algorithm has an approximation ratio of $1/4$ with $\beta=2$ for 
modality-based coverage, $\mu_M$, whereas a guarantee cannot be given for the 
other measures, $\mu_C$ and $\mu$.

Turning to the algorithm's complexity, we note that the decision whether to 
perform a swap needs $\mathcal{O}(|\mathcal{A}|)$ and, if so, the update takes 
$\mathcal{O}(k |\mathcal{A}|)$ time. Since the algorithm stores only the 
current solution, the space complexity is $\mathcal{O}(|V| + |E| + k(|V_q| + 
|E_q|))$.

A drawback of the swap-based algorithm is that it requires a static query, 
i.e., the rumour to explain is fixed. Otherwise, the relevance score of each 
subgraph is no longer valid. However, we later present a caching mechanism that 
enables the one-pass swap-based selection to be used also to explain different 
rumours.


\section{Graph Similarity with Embeddings}
\label{sec:rumour_embedding}

To avoid expensive graph computations when extracting an explanation for a 
rumour, this section shows how a novel technique for graph representation 
learning provides the basis for efficient example-based explanations. We first 
summarize how embeddings are incorporated in our approach 
(\autoref{sec:embedding_overview}), before introducing the representation 
learning technique for nodes (\autoref{sec:node_embedding}), subgraphs based on 
nodes (\autoref{sec:subgraph_embedding}), and subgraphs based on edges 
(\autoref{sec:edge_subgraph_embedding}).

%
%
%


\subsection{Overview}
\label{sec:embedding_overview}

\sstitle{Subgraph embedding}
A vector representation of a rumour is derived by subgraph embedding, 
as realised by an encoder and a decoder~\cite{duong2021efficient}. The 
former constructs a $d$-dimensional vector (where $d \ll 
|V|$), aka embedding, for a subgraph. The latter maps these 
vectors to a measure that reflects the similarity of two subgraphs. 
As such, the proximity of two subgraph embeddings in the vector space should 
capture the true similarity of two subgraphs. The difference between the 
proximity and the similarity is captured by a loss function, which is minimized 
when learning an embedding model by setting the parameters of the encoder and 
the decoder.

A model for rumour embedding that shall be applied for MSGs should capture the following aspects:
\begin{compactitem}
	\item \emph{Structure-preserving:} Embeddings of subgraphs shall capture 
	their structure, e.g., isomorphic subgraphs should be close. 
	\item \emph{Modality-aware:} The different types of nodes, i.e., modalities 
	shall be incorporated in the embeddings.
	\item \emph{Feature-aware:} The features assigned to nodes and edges shall 
	be incorporated in the embeddings. Note that traditional graph embeddings 
	commonly neglect edge features, as they are often observed in social 
	networks~\cite{tam2019anomaly}.
\end{compactitem}

\sstitle{Similarity computation}
The similarity of subgraphs is computed as the cosine similarity of their 
embeddings, normalized to $[0,1]$. Using the cosine 
similarity, we emphasize the immediate neighbourhood of the 
nodes, independent 
of their location in the graph~\cite{duong2021efficient}. 

\sstitle{Query Embedding}
The rumour to explain, $q$, is a subgraph over a multi-modal social 
graph $G$. To facilitate similarity search, an embedding of $q$ needs to 
be in the same space as the embedding of the MSG. 
This is achieved by using the model learned to embed the nodes and subgraphs of 
$G$ also for $q$. 
However, even in case the subgraph $q$ extends $G$, adding nodes and edges to 
it, the model is still applicable, since the model focuses 
on the neighbourhood structure rather than the exact location of the subgraphs. 

\subsection{Node Embedding}
\label{sec:node_embedding}

\sstitle{Graph convolutional network (GCN)}
Given the above requirements, a Graph Convolutional Network
(GCN)~\cite{dong2019multiple,song2022bi,song2021jkt,xue2022dynamic} provides a starting point 
for our embedding model. A GCN directly reflects the structure of 
a graph and the feature vectors assigned to nodes. A 
traditional GCN applies a $k$-hop aggregation, i.e., a function that 
incorporates the values of the neighbourhood of a node through recursive 
computation with a predefined depth $k$. Here, the initial embedding of the 
recursion is given by the node values (or initialized randomly or through a 
one-hot vector of its 
degree)~\cite{hamilton2017inductive}.

Yet, traditional GCNs ignore the types of nodes and $k$-hop 
aggregation would incorporate the complete neighbourhood of a node. 
Applying such a model to a MSG, thus, would mix up modalities.

\begin{figure}[!h]
	\centering
	\includegraphics[width=0.5\linewidth]{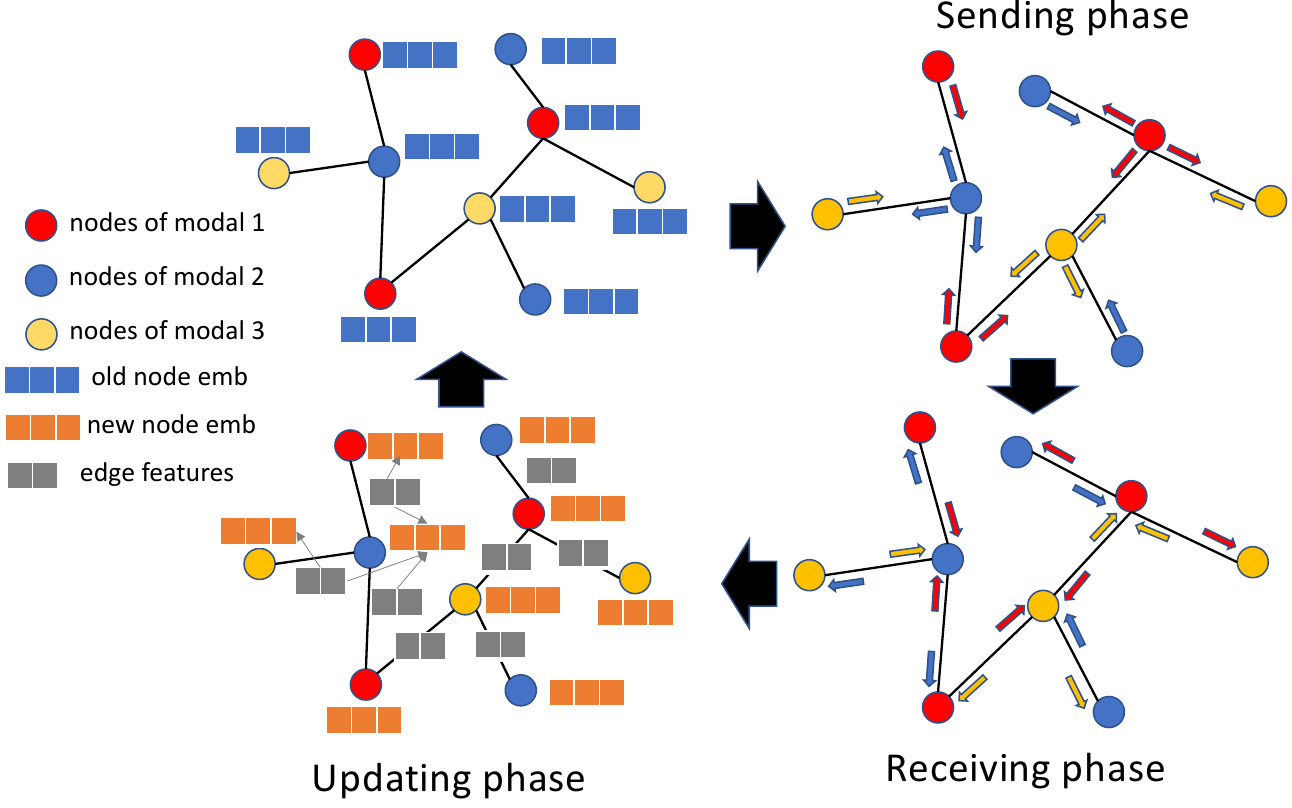}
	\caption{Message-passing graph embedding supporting modalities and 
	features. 
	}
	\vspace{-1em}
	\label{fig:message}
\end{figure}

\sstitle{Message-passing GCN (MPGCN)}
To overcome this limitation and achieve fine-granular control on the 
construction of an embedding, we incorporate a message-passing 
mechanism~\cite{duong2021efficient} in the GCN. A node representation is 
created by combining the representation of its own properties with those of its 
neighbours, through message-based interactions. A message sent from one node to 
its neighbours is constructed based on the node's current representation. Since 
messages are exchanged only between nodes connected by edges, the graph 
structure is incorporated. Each forward pass 
then includes three phases, also illustrated in \autoref{fig:message}: (i) a 
\emph{sending} phase, in which messages are broadcast to neighbours; (ii) a 
\emph{receiving} phase, in which messages from neighbours are consumed; and 
(iii) a \emph{updating} phase, which updates the embeddings.

\sititle{Sending}
Given a node $v$ at the $l$-th iteration, we send a message to its neighbours 
constructed from its current embedding $z^{(l)}_v$:
\[ m^{(l)}_{v \rightarrow u} = M^{(l)} z^{(l)}_v \]
where $M^{(l)}$ is a matrix that differs between iterations. 

\sititle{Receiving}
For a node $v$, the messages received from its neighbours are aggregated into a 
community embedding:
\[  z^{(l)}_{N(v)} = agg_l(\{m^{(l)}_{u \rightarrow v}, \forall\ u \in N(v)\}) \]
where $agg_l$ is an aggregation function and $N(v)$ is the set of neighbours of
$v$. Informally, the community embedding $z_{N(v)}$ reflects how node $v$ is 
related to its neighbours.

\sititle{Updating}
For a node $v$, a new embedding $z^{(l+1)}_{v}$ is computed using its 
embedding $z^{(l)}_v$ and its community embedding $z^{l}_{N(v)}$:
\begin{equation}
	\label{eq:update}
	z^{(l+1)}_{v} = combine_l(z^{(l)}_{N(v)}, z^{(l)}_v)
\end{equation} 
where $combine_l$ is a vector aggregation function that balances the intrinsic 
characteristics of a node and the influence of its neighbours.

Functions $agg_l$ and $combine_l$ are parametrized and may be changed in each 
iteration. In a traditional GCN, 
the aggregation is commonly defined as a component-wise maximum of the 
neighbouring embeddings after some linear 
transformation~\cite{hamilton2017inductive}:
\[agg_l(\{m^{(l)}_u, \forall \ u \in N(v)\})  = max(\{\sigma(W^{(l)}_{agg}
m^{(l)}_u + b^{(l)} ), \forall\ u \in N(v)\}).\]
Function $combine_l$ is typically a
simple concatenation of the neighbourhood embedding $z^{l}_{N(v)}$ and
the current embedding $z^{(l)}_v)$, before applying some non-linear
transformation:
\[combine_l(z^{(l)}_{N(v)}, z^{(l)}_v) = \sigma(W_{concat}^{(l)}
concat(z^{(l)}_{N(v)}, z^{(l)}_v)).\]
In this setting, $M^{(l)}, W^{(l)}_{agg}, b^{(l)}, W^{(l)}_{concat}$ are the 
parameters of the GCN that need to be learned.

\sstitle{Feature-aware MPGCN}
To incorporate features assigned to nodes and edges, we incorporate the 
following mechanisms:
\begin{compactitem}
	\item \emph{Node features:} We use node features in the message passing 
	 by initializing the first iteration with them:
	\begin{equation}
		z_v^{(0)} = f_{\phi(v)}(v)
	\end{equation}
	This way, both feature and structure characteristics of each node are 
	jointly propagated to compute embeddings. 
	\item \emph{Edge features:} We consider edge features in the updating step 
	of each iteration, as they act as a third type of information that is 
	embedded in the messages:
	\begin{equation}
		z^{(l+1)}_{v} = combine_l(z^{(l)}_{N(v)}, z^{(l)}_v, \mathit{concat}_{u \in N(v)}{f_{(\phi(u),\phi(v))}(u,v)})
	\end{equation}
	Here, we concatenate all features of associated edges of a node and reuse them for every iteration in order to regularise the messages by their intrinsic nature. 
\end{compactitem}

\sstitle{Modality-aware MPGCN}
The above GCN with message-passing enables us to incorporate types of nodes, 
i.e., modalities in 
the construction of embeddings, as required for MSGs. Next, we provide an 
instantiation of the above model, which we refer to as a heterogeneous MPGCN, 
or $h$-MPGCN for short. It adapts the message-passing procedure in the 
\emph{sending} 
and \emph{receiving} phases. 

\sititle{Sending}
The sent message now depends on the modalities $s = \phi(v)$ and $t = \phi(u)$ 
of the sending node $v$ and the receiving node $u$:
\begin{equation}
\label{eq:send}
  m^{(l)}_{v \rightarrow u} = M^{(l)}_{s,t} z^{(l)}_v
\end{equation}
where $M^{(l)}_{s,t}$ is a separate matrix for each pair of modalities $(s,t)$.

\sititle{Receiving} Instead of using the maximum as an aggregation function,
we sum up the messages from neighbours of a node $v$ to avoid the loss of the 
modality information.
\begin{equation}
\label{eq:agg}
  agg_l(\{m^{(l)}_{u \rightarrow v}, \forall u \in N(v)\}) = \sum_{u \in N(v)} \sigma(W^{(l)}_{agg} m^{(l)}_{u \rightarrow v} + b^{(l)})
\end{equation}
The idea being that neighbouring nodes potentially belong to different 
modalities, so that the sum retains information about all of them. An 
aggregation based on the maximum, as in a traditional GCN, in turn, would only 
keep the information of one modality. Moreover, it is known that an aggregation 
based on the maximum, in some cases, cannot distinguish between two
different neighbourhoods~\cite{renchi2020scale}.

\sstitle{Parameter training}
As common for neural networks, our model is trained by 
forward and backward propagation.

\emph{Forward propagation:} This phase operates like a message-passing 
algorithm with a pre-defined number of iterations. In each iteration, the 
parameters of the $h$-MPGCN, including 
$M^{(l)}_{s,t}$, $W^{(l)}_{\mathit{agg}}$, $b^{(l)}$, and $W^{(l)}_{\mathit{concat}}$,
are fixed; and the aforementioned steps of sending (\autoref{eq:send}), 
receiving (\autoref{eq:agg}), and updating (\autoref{eq:update}) are performed 
for each node. In brief, for each node $v$ in the MSG,
a message is sent to each of its neighbours $u$, which is constructed based on 
the modalities $\phi(v)$ and $\phi(u)$.
In the receiving step, given the messages that a node $v$ received from its
neighbours $I^{(l)}_v$, the function $agg_l$ is applied to obtain
the community embedding of node $v$, i.e., $z^{(l)}_{N(v)}$.
In the updating step, a new embedding of $v$ at iteration $l$+$1$ is obtained
by applying the function $combine_l$ to the community embedding 
$z^{(l)}_{N(v)}$ and its embedding of the previous iteration $z^{(l)}_v$.

\emph{Backward propagation:}
The model parameters (the message matrix $M$ and parameters of 
functions $agg$ and $combine$) are trained in an unsupervised manner. Initially, these parameters are assigned randomly, while, later, 
they are learned gradually using Stochastic Gradient Descent (SGD), taking into 
account a loss function that favours a small proximity of embeddings of 
neighbouring nodes:

\vspace{-.5em}
\smaller
\begin{equation}
\label{equ:loss_unsup}
L(z_v) = - Q^+E_{u_n \sim N_l(v)} \log(\sigma(z^T_v z_u)) - Q^-E_{u_n \sim \hat{N}_l(v)}\log(\sigma(-z^T_v 
z_{u_n}))
\end{equation}

\normalsize
\noindent
where $Q^+$ and $Q^-$ are the numbers of positive and negative samples, 
respectively; $N_l(v)$ is the set of all nodes from the one-hop neighbourhood 
to the $l$-hop neighbourhood of $v$ with the same modality; 
$\hat{N}_l(v)$ is the set of non-neighbouring nodes of $v$ with a 
different modality. As such, we randomly select $Q^+$ and $Q^-$ 
neighbouring and non-neighbouring nodes of $v$, respecting its modality.

The time and space complexities of one forward/backward propagation are 
$\mathcal{O}(|V| deg (G))$ and $\mathcal{O}(deg(G))$, respectively, where 
$deg(G)=\max\{deg(v)|v \in V_G\}$ is the maximum degree of $G$.

\subsection{Node-based Subgraph Embedding}
\label{sec:subgraph_embedding}


Our approach to embed subgraphs of a MSG $G$ exploits the above model that 
returns embeddings for all nodes. This model, learned on the whole graph, 
captures the graph's structure in a comprehensive manner. Hence, for a subgraph 
$H$ of $G$, we can project the model on the respective nodes. Such a projection 
is akin to truncated message passing, in which solely the nodes in $H$ send 
messages to neighbouring nodes that are also in $H$. 
Note though that the 
parameters of the functions used for sending, receiving, and updating are 
taken from the $h$-MPGCN learned to embed individual nodes.


The above process yields embeddings for all nodes of a subgraph. Since each 
embedding summarises the node's receptive field, i.e., the subgraph, it is a 
candidate to represent the whole subgraph. Motivated by this observation, we 
follow a compositional approach and compute the average of the node embeddings 
to obtain an embedding for the whole subgraph.

\subsection{Edge-based Subgraph Embedding}
\label{sec:edge_subgraph_embedding}

Finally, we consider an adapted approach to obtain subgraph embeddings that 
incorporate solely the edges of a multi-modal social graph. The idea here being 
that edges can be considered the smallest possible subgraphs, so that edge 
embeddings may be more discriminative than node embeddings. 
To realize this idea, we first construct edge embeddings by averaging the 
embeddings of the adjacent nodes. Second, we again adopt a compositional 
approach and average the edge embeddings to represent the subgraph. This is 
equivalent to a degree-weighted combination of the respective node embeddings, 
as follows:
\begin{equation}
z_{s} = \frac{1}{|E_s|} \sum_{ (u,v) \in E_s} z_{(u,v)} 
= \frac{1}{|E_s|}\sum_{(u,v) \in E_s} \frac{z_u + z_v}{2} 
= \frac{1}{2|E_s|} \sum_{u\in V_s}deg(u)z_u
\end{equation}
In other words, an edge-based subgraph embedding is similar to information 
diffusion: The influence of a node is amplified by its degree and the influence 
of a rumour on entities and relations in a social graph is normalised by the 
number of its propagations $|E_s|$. It is noteworthy there are many other similarities. For example, soft cosine similarity is a measure that originally computes the similarity between two documents even though they have no words in common~\cite{adhikari2018sub2vec}. The idea is to convert the words into respective word vectors, and then, compute the similarities between them. In the context of graph, we can design a graph-based soft cosine similarity between two subgraphs by computing the similarities between their respective nodes. However, soft cosine similarity or any other similarities such as average similarity both need an aggregation of individual similarity to compute group similarity (e.g. a document vector is a sum of belonging word vectors). Our approach is a generic method that computes a weighted aggregation of node embeddings into the subgraph embedding. Here, the weights are designed as the respective degree of the nodes so that their importance are reflected during similarity computation.

\section{Optimisations for Efficiency and Robustness}
\label{sec:extension}


Finally, we show how to improve the efficiency and robustness of our approach, 
by means of indexing (\autoref{sec:indexing}), caching (\autoref{sec:caching}), 
and handling of concept drift (\autoref{sec:drift}).

\subsection{Indexing}
\label{sec:indexing}

In our context, the relative similarity of embeddings is more important than
their absolute similarity. Since we approach the problem of subgraph selection 
based on embeddings, we can use nearest neighbour
search in the embedding space. This way, we can relay on a large body of work 
on indexing for fast nearest neighbour
search in numeric spaces, including R-trees~\cite{guttman1984r}
and kd-trees~\cite{de2008orthogonal}.

Note that these indexing techniques can be applied to assess the cosine
similarity by normalizing each embedding to have a length of one. In this case,
the cosine similarity corresponds to the dot product between two embeddings,
which is negatively correlated with their Euclidean distance.
Moreover, several variants of R-trees and kd-trees to handle high-dimensional 
embeddings have been
proposed~\cite{stepivsnik2021oblique}. In our
experiments, we later adopt an improved version of
the kd-tree~\cite{stepivsnik2021oblique}.

\subsection{Caching}
\label{sec:caching}

The subgraph of the rumour to explain may itself be part of explanations for 
other rumours. 
This observation motivates the design of a cache for subgraphs that are 
frequently used in explanations. Specifically, our idea is based on the 
incremental k-medians clustering algorithm. Each time a new rumour is detected, 
we update the $k$ medians. For each median, we maintain an explanation using 
the one-pass swap-based selection algorithm in 
\autoref{sec:construct_explanation}, where the relevance scores of subgraphs 
are evaluated by the new median.

Using such a caching mechanism, the one-pass selection algorithm is 
tailored to explain multiple rumours by maintaining a set of candidate 
subgraphs for each median rumour. However, when a median changes, we need 
to recompute the candidate set, via greedy selection or 
the basic swap-based algorithm that includes multiple passes over the data. 
Hence, the one-pass selection strategy shall be used solely when the clustering 
of rumours is stable.

%

 \subsection{Dealing with Concept Drift}
 \label{sec:drift}

As a multi-modal social graph is extended over time with new nodes and edges, 
new neighbourhood structures may appear in the graph, potentially resulting in 
a concept drift. Another type of concept drift is that new rumour patterns can emerge and be different from past rumour examples. Such drifts can alter the level of 
similarity of neighbouring or non-neighbouring nodes are supposed to have.

\sstitle{Embedding update}
Our message-passing mechanism enables us to handle such phenomenon by 
performing a few iterations in the parameter training process. As a result, 
convergence is reached faster than by completely retraining the model as it 
would need to be done for traditional GCNs. 

The question is how frequent such adaption shall be incorporated, though. A 
straightforward approach is to perform the adaption after a fixed number of 
entities or relations have been added to the graph. However, such approach may 
incur unnecessary overhead if there is no concept drift. However, we observe 
that concept drift will cause a change in the assessment of dissimilar 
subgraphs, which can be exploited for drift detection, as follows.

\sstitle{Drift detection}
We maintain a set of anchor subgraphs $R=\{r_1, \ldots, r_k\}$ of $G$ (e.g., 
using the $k$ medians of the incremental $k$-medians clustering in 
\autoref{sec:caching}). Then, we index a rumour subgraph $s \in \mathcal{S}$ by 
computing the dissimilarity between the embeddings of $s$ and the anchor graphs 
$R$:
\[
\zeta_s = [dist(s,r_1), \ldots, dist(s,r_k)].
\]
The vector $\zeta_s$ is referred to as the dissimilarity representation of $s$. 
If there is no change in characteristics of the underlying graph, there is no 
change in the dissimilarity space. That is, any new 
rumour subgraphs would be generated according to a stationary probability 
distribution and the dissimilarity vectors would follow a stationary 
probability distribution. Hence, when a concept drift happens, the new rumour 
subgraphs would no longer be generated according to the old stationary 
distribution. Then, the change in the distributions can be identified using the 
CUSUM test, see~\cite{zheng2021semi}, which is based on the Central Limit 
Theorem for multivariate vector streams.

%


\section{Performance Evaluation}
\label{sec:exp}

We evaluated our approach in various experimental settings, as detailed in 
\autoref{sec:setup}. Below, we report on the effectiveness 
(\autoref{sec:effectiveness}) and efficiency (\autoref{sec:efficiency}) of 
example-based explanations; their utility characteristics 
(\autoref{sec:explainability}), properties of the employed embeddings 
(\autoref{sec:correctness} and \autoref{sec:efficiency_emb}), before concluding 
with a qualitative discussion (\autoref{sec:quality}) and experiments on 
adaptivity (\autoref{sec:adaptivity}).

\subsection{Experimental Setup}
\label{sec:setup}

\sstitle{Datasets}
We used several benchmarks in rumour detection:

	\emph{Snopes:} This dataset\footnote{\url{http://resources.mpi-inf.mpg.de/impact/web_credibility_analysis/Snopes.tar.gz}} originates from the {by far 
	most reliable 
and 
largest platform for fact checking~\cite{vosoughi2018spread}}, covering different domains such as news websites, social media, and e-mails. 
The dataset comprises 80421
documents of 23260 sources that contain 4856 labelled stories (as rumour or non-rumour).
The multi-modal graph has been constructed by taking unique, curated stories from Snopes and using them as a query for a search engine to collect Web pages as documents, while the originating domain names indicate the sources. The top-30 retrieved documents are linked to a given story, except those that originate from Snopes in order to avoid a bias~\cite{popat2017truth}.

	\emph{Anomaly:} This dataset contains 4 million 
	tweets, 3 million users, 28893 hashtags, and 305115 linked 
	articles~\cite{tam2019anomaly}. The data spans over different domains, such 
	as politics, 
	fraud \& scam, 
	and crime. 
	
	\emph{COVID-19 news:} We assembled a collection of datasets 
	related to COVID-19 from Aylien~\cite{montariol2021scalable}, 
	AAAI\footnote{\url{https://github.com/diptamath/covid_fake_news/}}, and further sources\footnote{\url{https://data.mendeley.com/datasets/zwfdmp5syg/1}}$^{,}$\footnote{\url{https://doi.org/10.5281/zenodo.4282522}}. For example, 
	the Aylien data covers 1.7 million global news,  215657 hashtags, 136847 
	users, and 440 news sources throughout the COVID-19 outbreak (Nov 2019 - 
	July 2020)~\cite{montariol2021scalable}.

%

\sstitle{Synthetic data}
To generate synthetic data for controlled experiments, we used several 
propagation models. Such models can be divided as follows~\cite{dong2019multiple}. 

	\emph{Infection model:} This class of models was originally proposed to 
	describe the spread of diseases among individuals. Widely used 
	infection models are the SI (Susceptible-Infected) model, the SIS 
	(Susceptible-Infected- Susceptible) model, and the SIR 
	(Susceptible-Infected-Recovery) model. Individuals in these models 
	assume three states, Suspected, Infected, and Recovery, and may change 
	between some of these states with a certain probability. For example, in 
	the SI model, a suspected individual becomes infected with a certain 
	probability. 
	Due to the similarity of the propagation of diseases and rumours, 
	infection models are 
	widely adopted for social media~\cite{wang2014rumor}.
	
	\emph{Influence model:}  An influence model focuses on 
	the effects of neighbours on one's decision, with IC (Independent Cascade) 
	and LT
(Linear Threshold) being two widely-adopted models. The IC model is an 
iterative model, where in each iteration, each node has a chance to activate 
its neighbours. If a neighbour $v$ is not activated by the node $u$ 
successfully, it will never be activated by $u$ again. However, $v$ can still 
be activated by other neighbours. This iteration is repeated until the model 
converges. In the LT model, each node is initialized with a threshold $\theta$, 
which describes the probability of the node to be infected. Each edge $(u,v)$ 
is initialized with a weight $w(u,v)$.  In each iteration, for a node $v$, the 
total influence of its neighbours $N(v)$ will be summarized with $I(v) = 
\sum_{u \in N(v)} w(u,v) I(u)$. If $I(v) > \theta$, the node $v$ will be 
activated. The iteration is repeated until convergence is reached.

In this work, we rely on data generated based on these infection models and 
influence models, as ground truth as well as training data. The features are taken from the real datasets and assign to nodes randomly. Details about this 
generation can be found in~\cite{dong2019multiple}.

\sstitle{Baselines} To construct numeric representations of graphs, we compare 
with the following baseline techniques:
\begin{compactitem}
	\item \emph{GGSX~\cite{ggsx}:} is a structure-based graph index that uses 
	paths with bounded length as features.
	\item \emph{CTIndex~\cite{ctindex}:} is a structure-based graph index that 
	identifies both paths and cycles of interest to create graph fingerprints. 
	\item \emph{Sub2Vec:} An embedding technique for subgraphs based on 
	neighbourhood information~\cite{adhikari2018sub2vec}. Similar to our 
	edge-based version of subgraph embedding, \emph{Sub2Vec} also leverages the 
	information about node degrees and subgraph sizes in the embedding. However 
	\emph{Sub2Vec} lacks any mechanisms for the efficient construction of 
	embeddings in dynamic settings.
\end{compactitem}
Turning to the similarity of graphs, we compare our approach based on 
embeddings against the following measures (see also \autoref{sec:similarity}):
\begin{compactitem}
	\item \emph{MCS~\cite{zhu2013high}:} counts the number of nodes and edges in the maximum 
	common subgraph of two graphs. 
	\item \emph{Graphsim~\cite{yu2014subgraph}:} extends the MCS measure by 
	incorporating modalities of the data. 
	\item \emph{Graph edit distance (GED):} lifts the idea of the string edit 
	distance to graphs~\cite{duong2021efficient}. 
\end{compactitem}

\sstitle{Metrics} We use the following measures:

\emph{Explanation time:} We assess the total runtime of constructing an 
explanation for a rumour. 

\emph{Runtime:} We evaluate the runtime of various operations in our framework, 
including the time needed for embedding the graph, for embedding the rumours, 
and for computing the subgraph similarity. 

\emph{F1-score:} A good explanation shall be useful also for rumour 
detection. Hence, we evaluate the predictive power of an explanation by 
incorporating it in a classifier. Specifically, we design a classifier based on 
a graph neural network with an attention mechanism~\cite{xuan2022dynamic}. It 
computes an attention map based on the maximum similarity between the input 
subgraph and each of the explaining examples and aggregates these attention 
maps with a multi-layer architecture~\cite{zhang2019multi}. In other words, the training process is basically the same. The only difference is that now the selected examples are considered as prototypes where the GNN pays additionally attentions to these prototypes in a middle prototypical layer. We then measure the 
F1-score of this classifier.

\emph{Memory cost:} 
We evaluate how much space is needed to store a rumour subgraph upon arrival 
(only for the explanation purpose).

\sstitle{Environment}
Our results have been obtained on a workstation with an AMD Ryzen Threadripper 1900X 8-Core CPU @ 3.8GHz with 62 GB RAM and an Nvidia GTX 1080Ti GPU with 12 GB.
We report average results over 10 experimental runs. Unless stated otherwise, 
we use a query size of 10, a cache size of 20, and an embedding size of eight.

\subsection{Effectiveness of Explanations}
\label{sec:effectiveness}

We first assess the predictive aspect of our approach through the accuracy of a 
trained classifier, as introduced above.


\begin{figure}[!h]
\vspace{-.5em}
  \centering
  \begin{minipage}{0.47\linewidth}
    \centering
  	\includegraphics[width=0.7\linewidth]{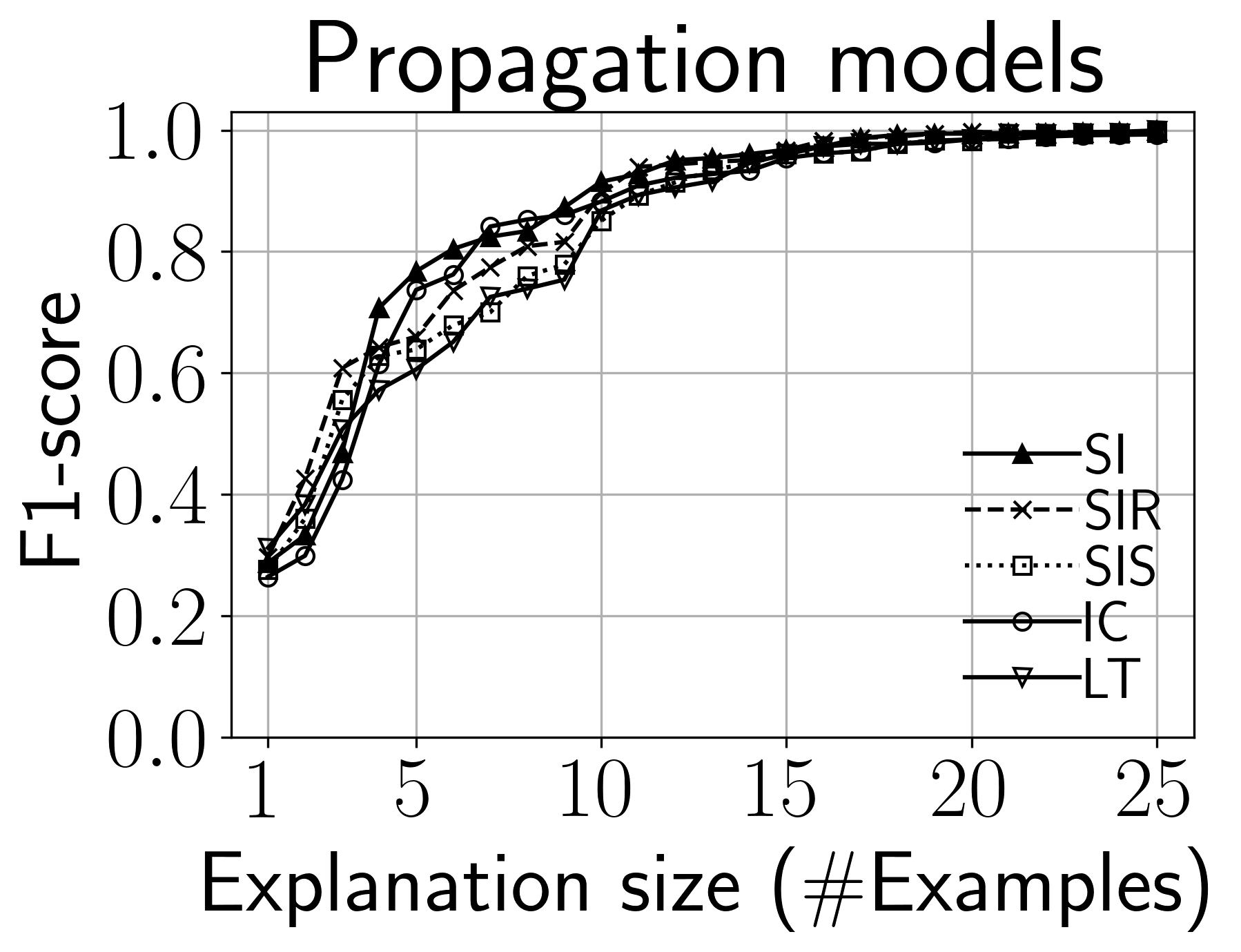}
\vspace{-1em}
  \caption{Effectiveness for synthetic data}
  \label{fig:effectiveness_synthetic}
  \end{minipage}
  \quad
  \begin{minipage}{.47\linewidth}
\centering
  \includegraphics[width=0.7\linewidth]{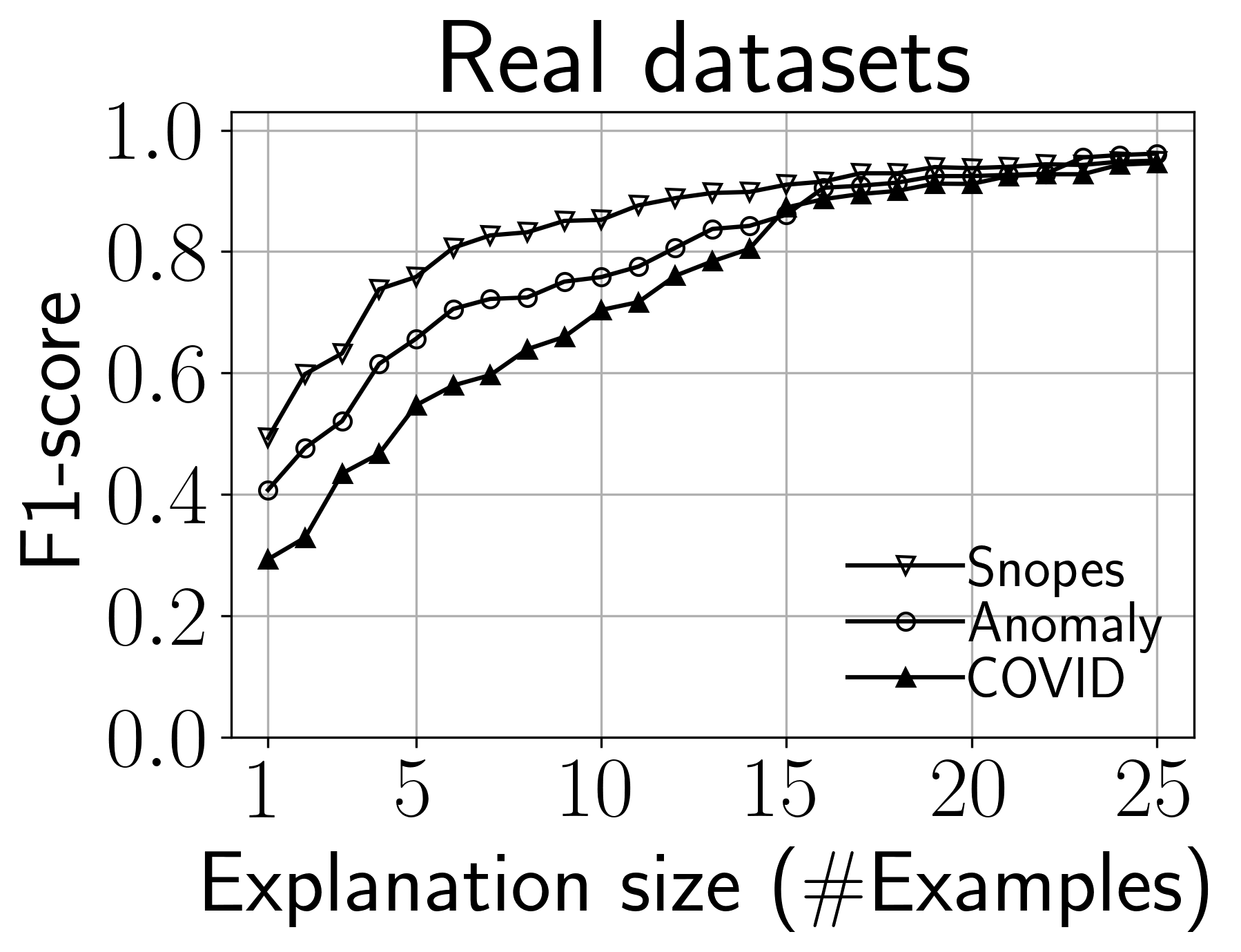}
\vspace{-1em}
  \caption{Effectiveness for real-world data}
  \label{fig:effectiveness_real}
  \end{minipage}
\vspace{-1em}
  \end{figure}

\sstitle{Propagation models}
Using synthetic data, we investigate the robustness of our explanations for 
different propagation models. 
\autoref{fig:effectiveness_synthetic} shows the F1-score of the explanation 
relative to the explanation size, taking the top-$k$ examples in the 
explanation, varying $k$ from 1 to 25. Initially, the F1-score is small 
due to the lack of information in a small explanation for detection purposes. 
However, when the explanation size increases, the predictive power increases 
quickly. Also, when the explanation size is large enough ($\geq 10$), 
the F1-score converges to similar values for all propagation models, 
highlighting the robustness of our model. 

\sstitle{Real-world datasets}
Adopting the same setup for real-world datasets, a similar trend can be 
observed in \autoref{fig:effectiveness_real}. The F1-score increases when the 
explanation size increase because there is more information in the explanation 
for rumour detection. Interestingly, the convergence is slower than for the 
synthetic data. A reason could be that social media includes different rumour 
structures, so that more information is needed to capture them. 


\subsection{Efficiency of Explanations}
\label{sec:efficiency}

We evaluate the efficiency of our framework by measuring the total 
runtime when user asks for an explanation. To this end, we choose a rumour 
randomly as a query and consider all rumours as a historic data. We do so for 
each rumour in each dataset and report the averaged result. 

\begin{figure*}[!h]
\centering
     \begin{subfigure}[b]{0.31\textwidth}
         \centering
         \includegraphics[scale=0.3]{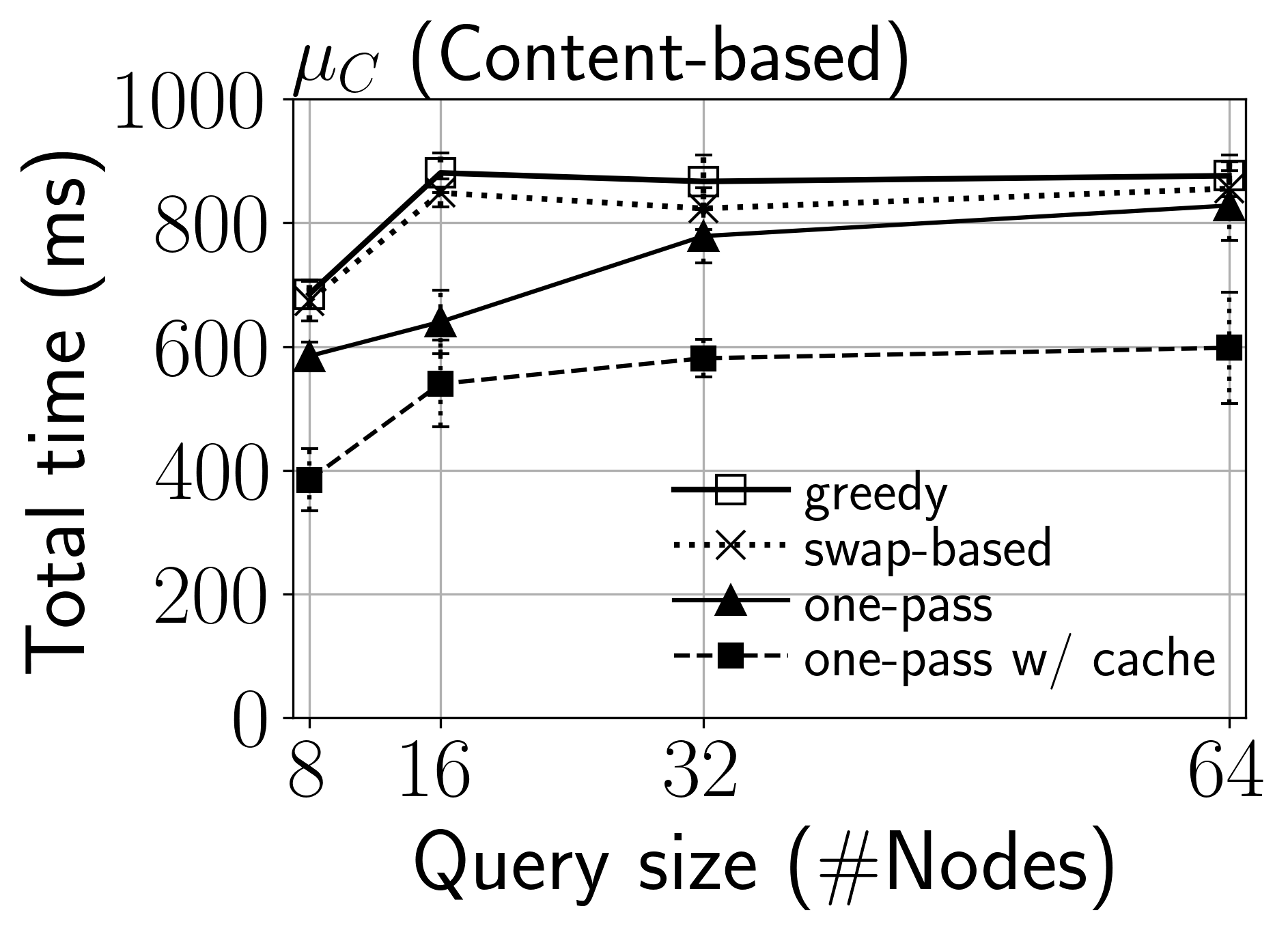}
         \caption{Content-based}
         \label{fig:efficiency_f1}
     \end{subfigure}
     \begin{subfigure}[b]{0.31\textwidth}
         \centering
         \includegraphics[scale=0.3]{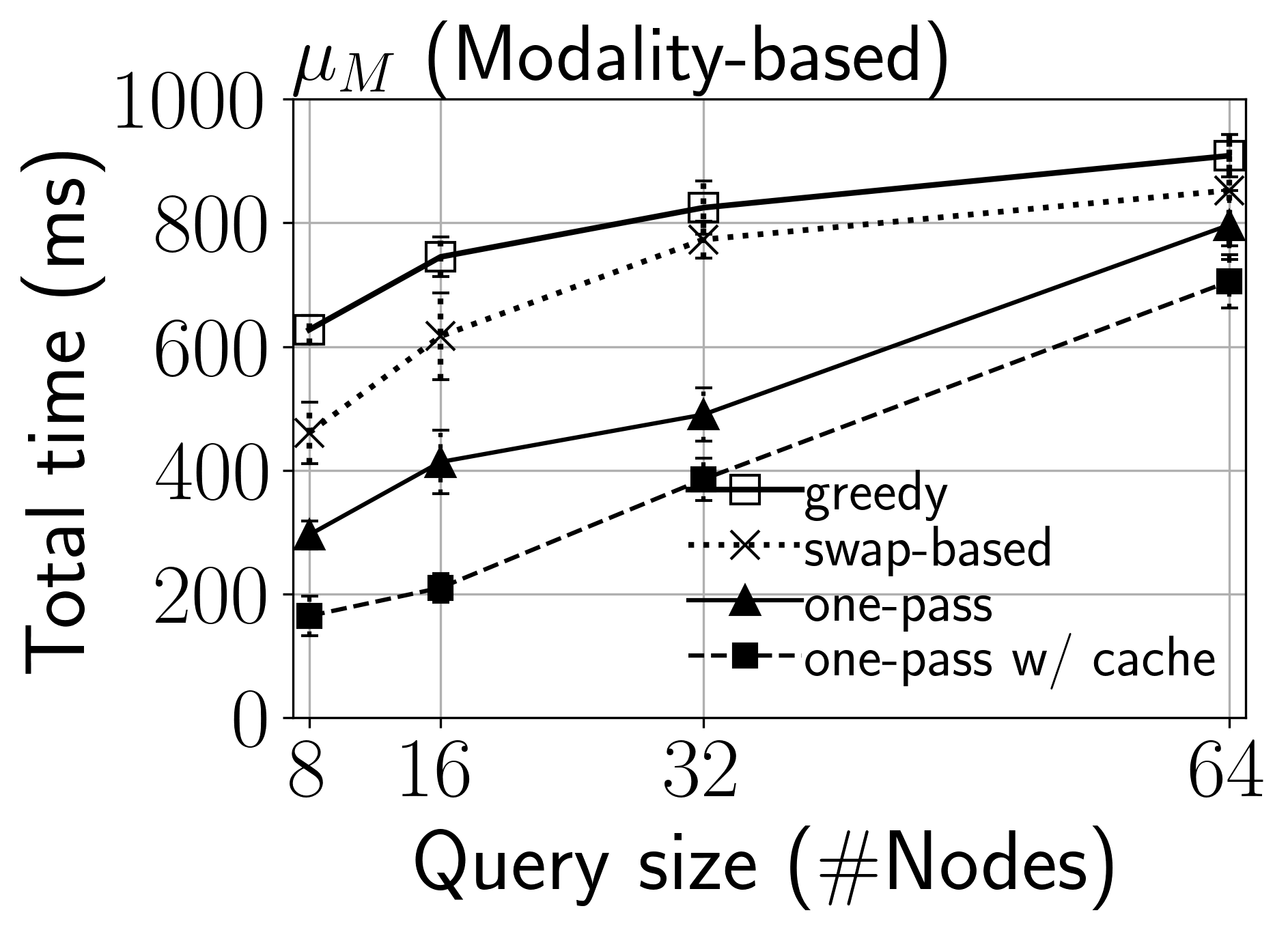}
         \caption{Modality-based}
         \label{fig:efficiency_f2}
     \end{subfigure}
     \begin{subfigure}[b]{0.31\textwidth}
         \centering
         \includegraphics[scale=0.3]{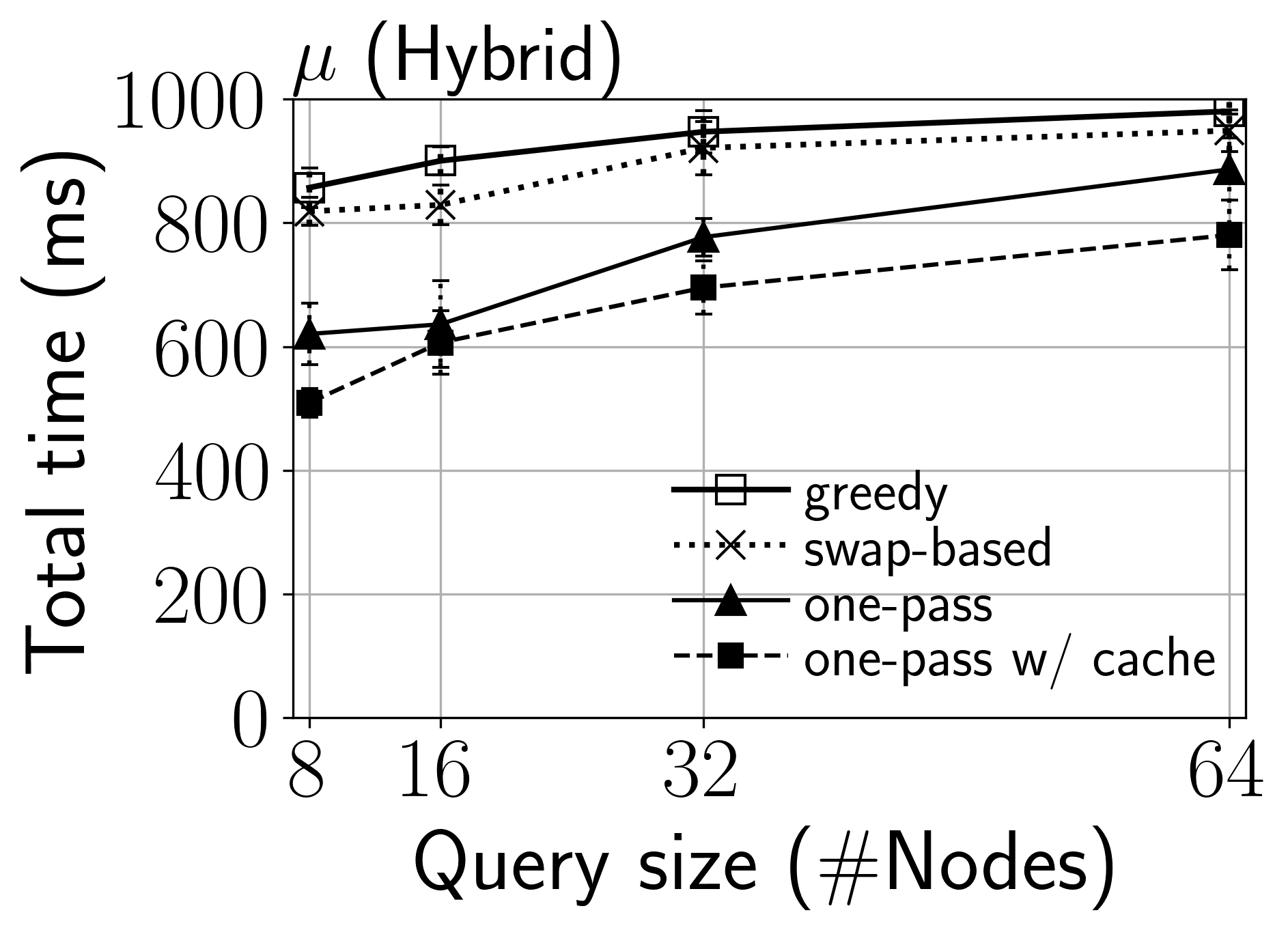}
         \caption{Hybrid}
         \label{fig:efficiency_f}
     \end{subfigure}
\vspace{-.5em}
        \caption{Effects of Query Size on Explanation Efficiency}
        \label{fig:efficiency}
\vspace{-.5em}
 \end{figure*}

\sstitle{Effects of query size}
We vary the number of nodes in the rumour used as a query, while fixing 
the explanation size to 15. To have a consistent query size, we 
add dummy nodes to a rumour until a pre-defined size is reached 
(8, 16, 32, 64). The dummy nodes are added randomly to the neighborhood of nodes in the rumor. The features of dummy nodes are instantiated to a random value. We also study different utility functions ($\mu_C$, $\mu_M$, 
$\mu$, see \autoref{sec:goodness}). \autoref{fig:efficiency} shows that the 
explanation time is small, less than $1s$ in all cases. 
The explanation time increases linearly with the explanation size, 
since larger queries need more time for embedding. 
Among different selection strategies, as expected, the one-pass algorithm with 
caching mechanism is the fastest. 
In some cases, the explanation time for smaller queries ($k=16$ for $\mu_C$ 
with greedy and swap-based selections) can be higher than for larger queries. 
This is because there could be more similar rumours with smaller sizes to 
consider for explanation, depending on the size distribution of rumours in the 
datasets.

\begin{figure*}[!h]
\centering
     \begin{subfigure}[b]{0.31\textwidth}
         \centering
         \includegraphics[scale=0.3]{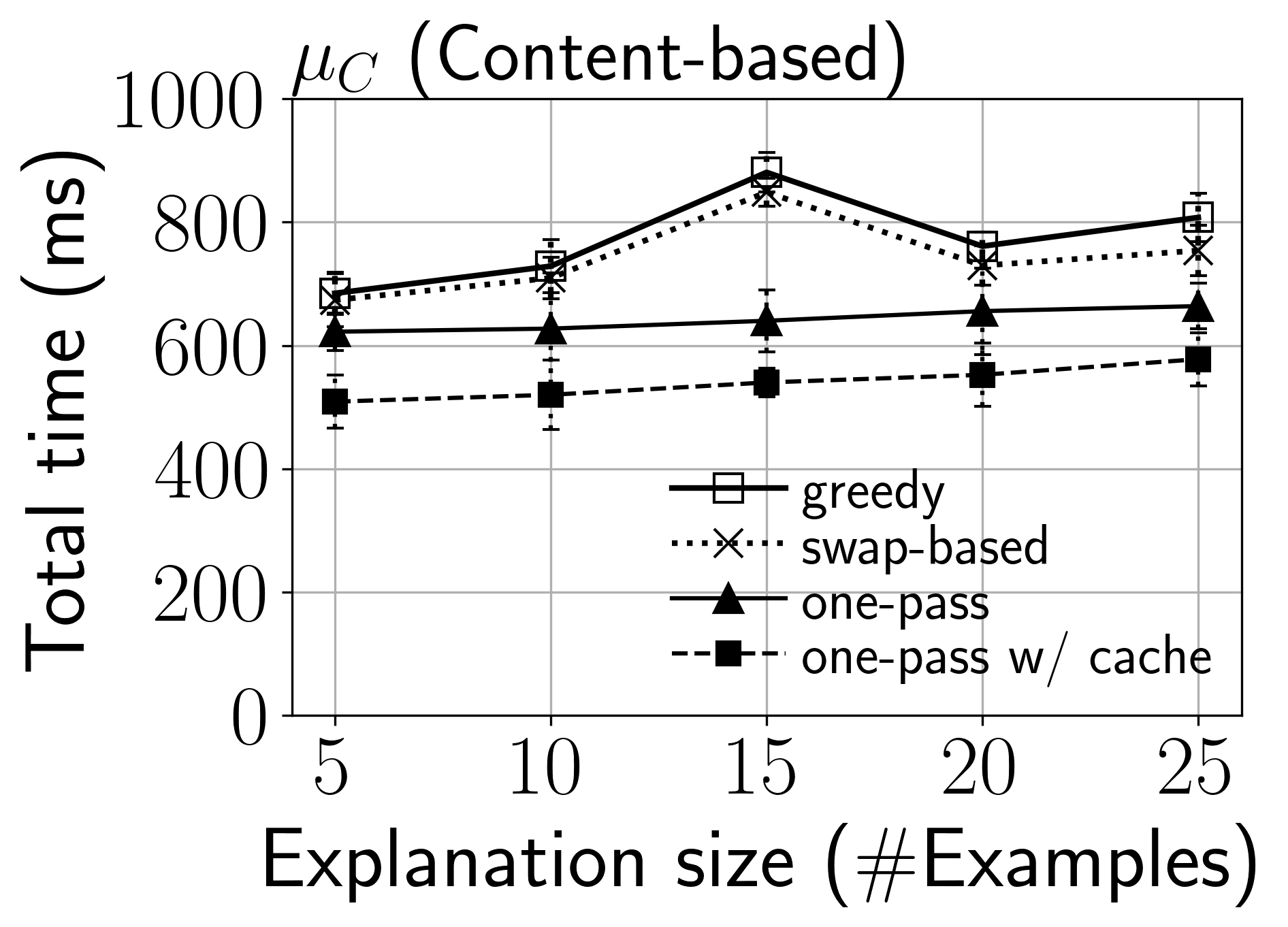}
         \caption{Content-based}
         \label{fig:efficiency_f1_k}
     \end{subfigure}
     \begin{subfigure}[b]{0.31\textwidth}
         \centering
         \includegraphics[scale=0.3]{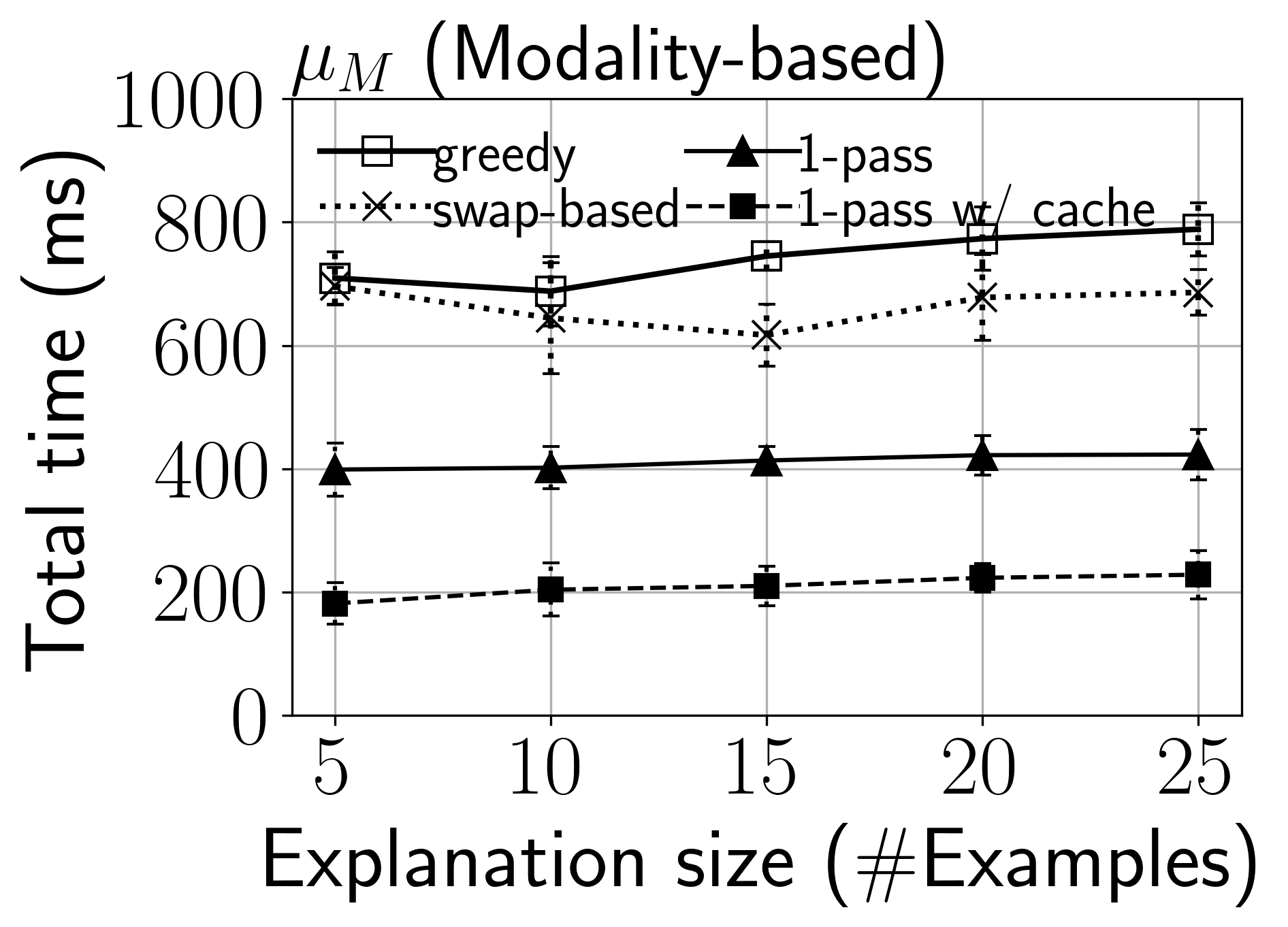}
         \caption{Modality-based}
         \label{fig:efficiency_f2_k}
     \end{subfigure}
     \begin{subfigure}[b]{0.31\textwidth}
         \centering
         \includegraphics[scale=0.3]{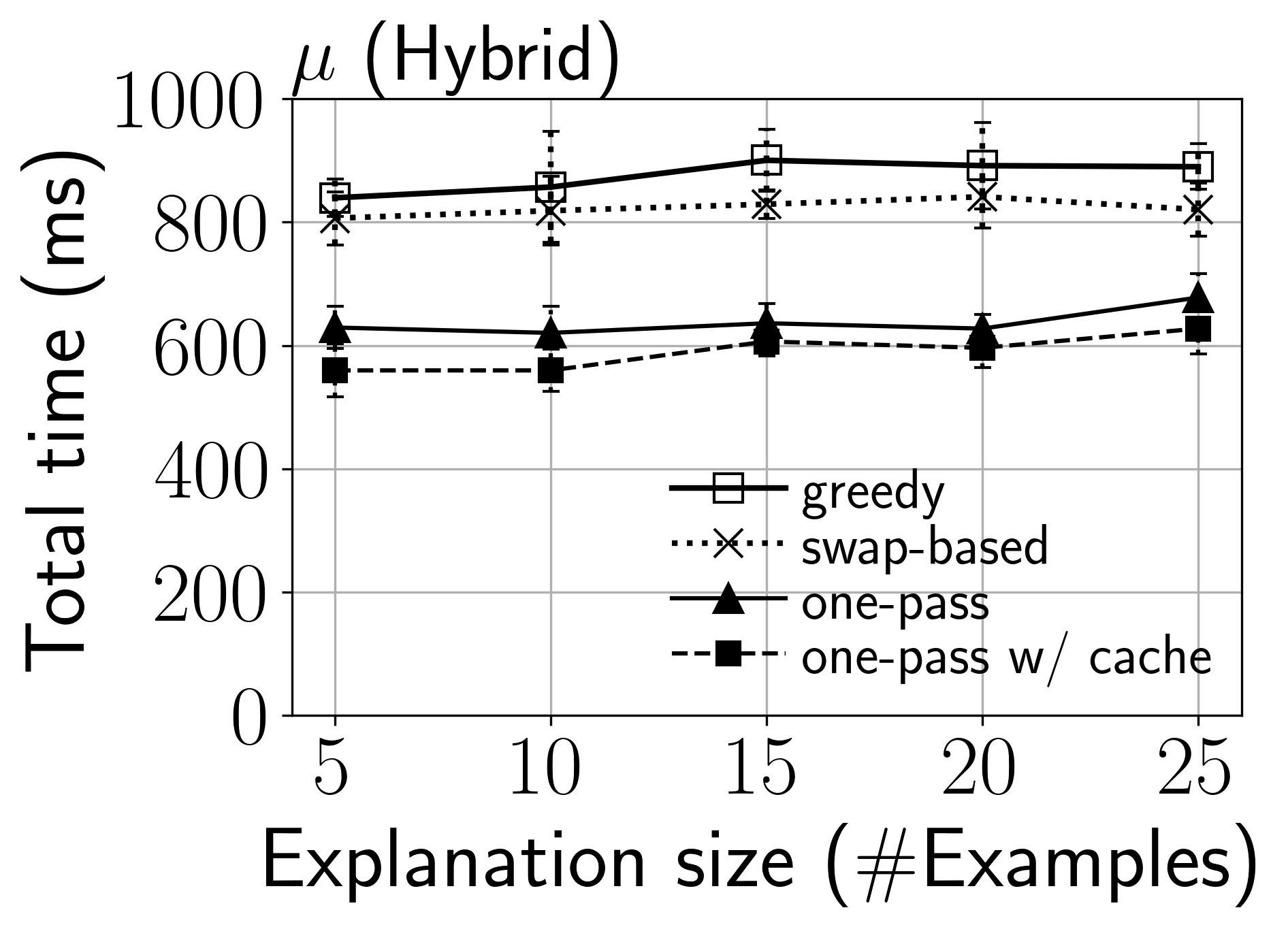}
         \caption{Hybrid}
         \label{fig:efficiency_f_k}
     \end{subfigure}
\vspace{-.5em}
        \caption{Effects of Explanation Size on Explanation Efficiency}
        \label{fig:efficiency_k}
\vspace{-.5em}
 \end{figure*}

\sstitle{Effects of explanation size}
Here, we vary the number of examples in the explanation from $k=5$ to 
 $k=25$, while fixing the query size to 16. 
 Again, we consider the different utility functions. As seen in   
 \autoref{fig:efficiency_k}, the 
 explanation time is larger for bigger $k$, but the increase is not drastic. 
 Considering that 15 elements may typically be handled by users in terms of 
 cognitive load~\cite{wang2021cognitive}, 
 the explanation can be retrieved with less than $1s$. 
%

\subsection{Utility of Explanations}
\label{sec:explainability}

Next, we investigate the utility of graph-based explanations. Since users in 
different application domains can have different perspectives on 
utility~\cite{gan2018extracting}, we follow a utility-based approach of evaluating explanations~\cite{zhao2021scalable,nimmy2022explainability}. In particular, we evaluate the utility difference (increase or decrease) of explanations derived with different selection algorithms.

\begin{figure}[!h]
  \centering
  \begin{minipage}{0.47\linewidth}
    \centering
  	\includegraphics[width=0.7\linewidth]{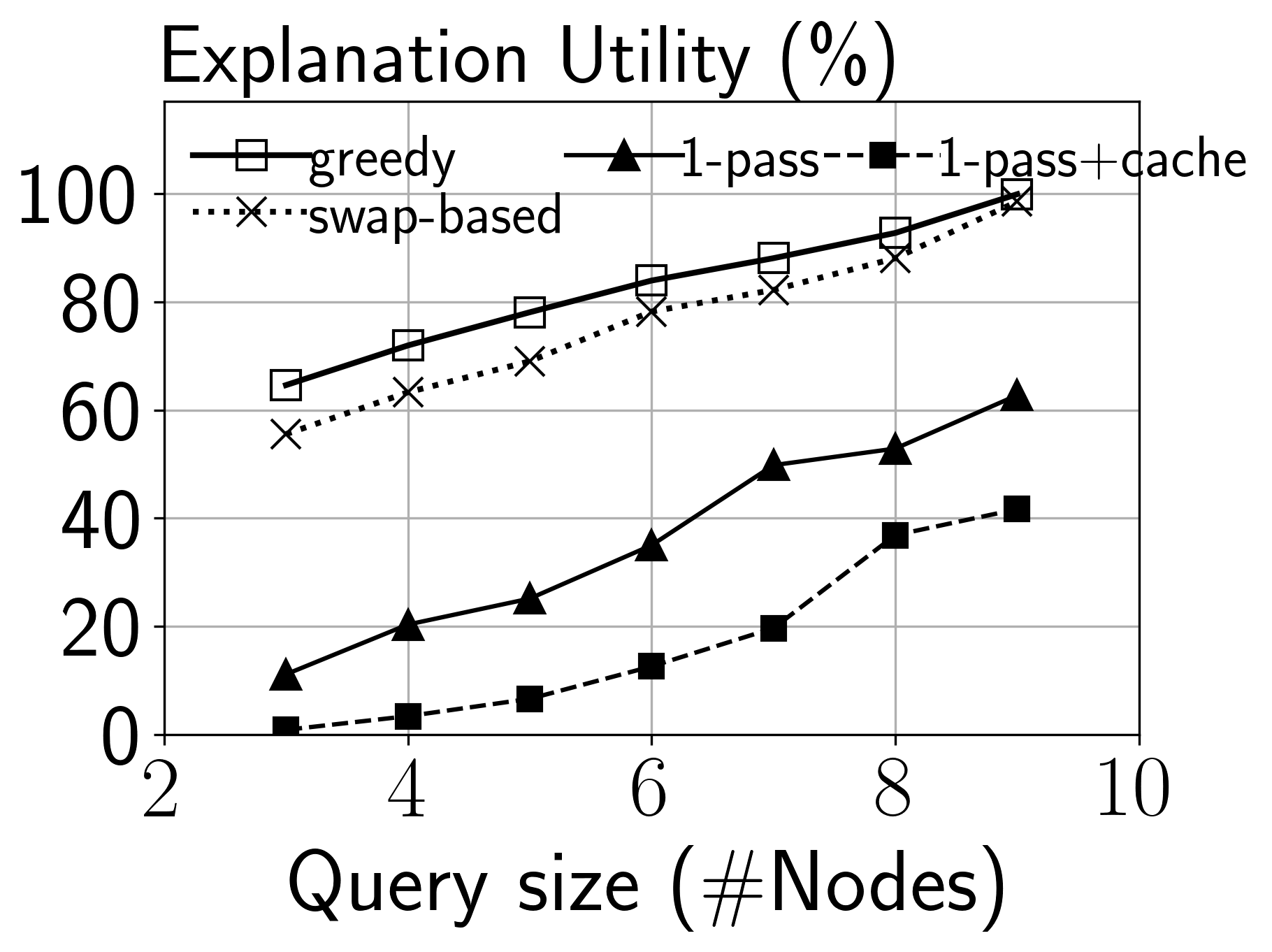}
  \caption{Explanation quality by query size}
  \label{fig:explainability_query}
  \end{minipage}
  \quad
  \begin{minipage}{.47\linewidth}
\centering
  \includegraphics[width=0.7\linewidth]{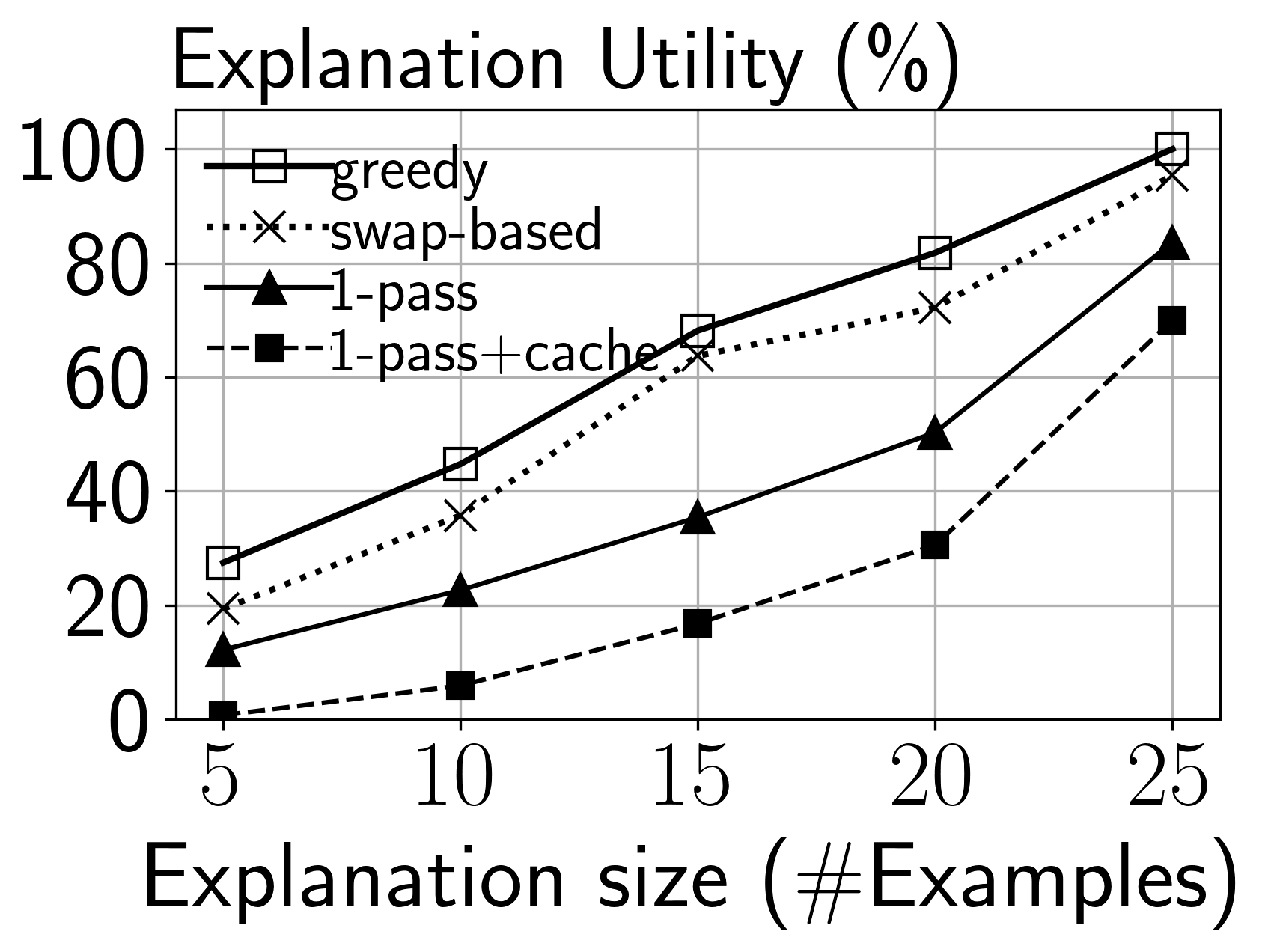}
  \caption{Explanation quality by explanation size}
  \label{fig:explainability_k}
  \end{minipage}
  \end{figure}

\sstitle{Effects of query size}
Similar to the above setting, we vary the number of nodes in 
the rumour used as a query from three to 
nine, and report average results. 
\autoref{fig:explainability_query} shows the utility 
(\autoref{sec:goodness}) scaled into the percentage of the maximum value. The 
utility
increases when the query size increases, since the selected subgraphs 
tend to be larger (to be similar with the query). The \emph{greedy} selection 
strategy yields the best results, 
achieving more than 60\% utility with a query size of only three. 

\sstitle{Effects of explanation size}
We also study the effects of the number of examples on the utility. Again, we 
scale the utility based on the maximum value. 
\autoref{fig:explainability_k} indicates that the utility increases with the 
explanation size, as more examples usually provide more information. The 
\emph{greedy} selection strategy performs best. Yet, this time, it 
achieves less than 30\% of utility when the explanation size is five. This 
highlights that it is difficult to provide a good explanation when the number 
of allowed examples is small.

\begin{figure}[!h]
\centering
     \begin{subfigure}[b]{0.47\linewidth}
         \centering
         \includegraphics[width=0.7\linewidth]{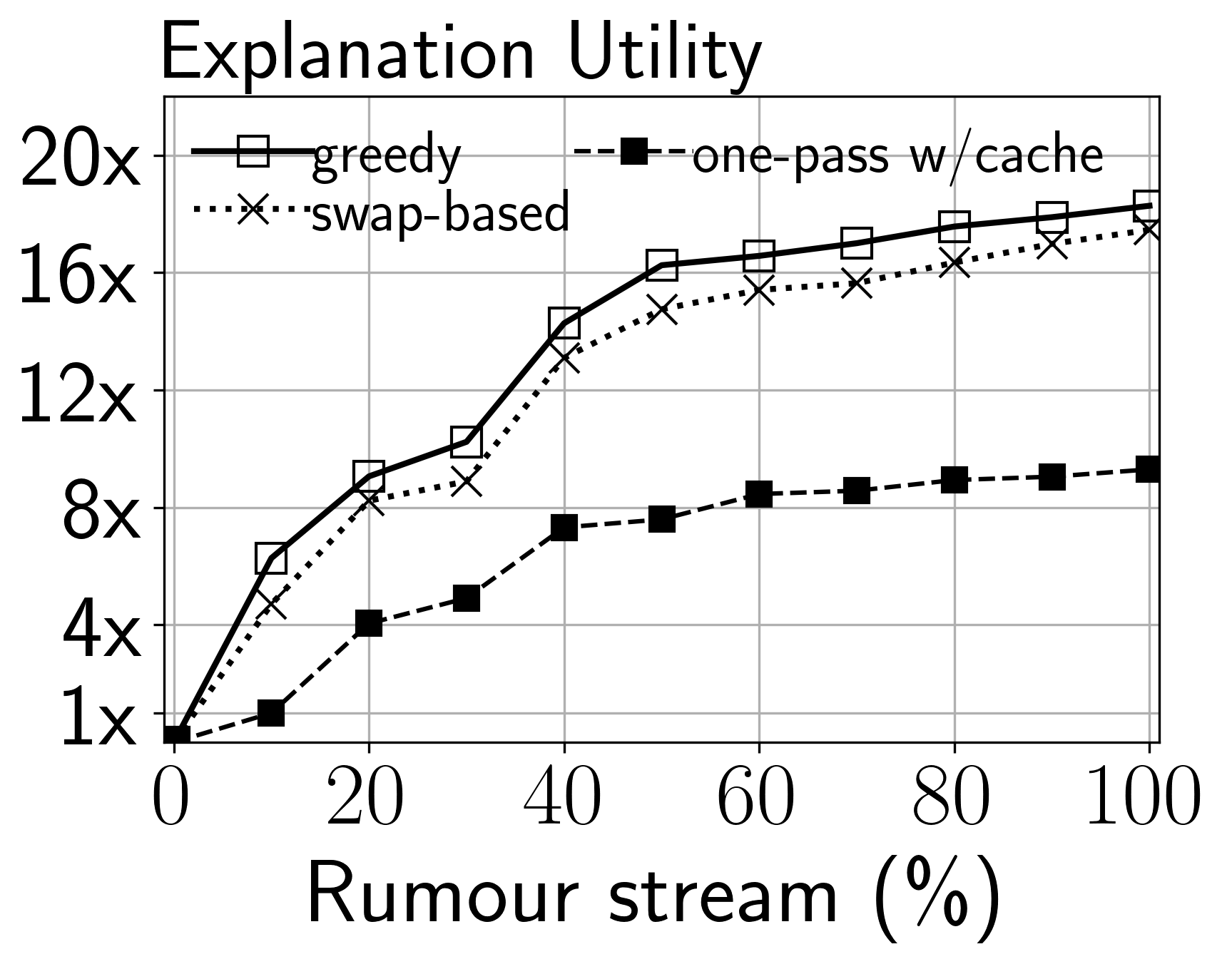}
         \caption{Snopes dataset}
     \end{subfigure}
     \begin{subfigure}[b]{0.47\linewidth}
         \centering
         \includegraphics[width=0.7\linewidth]{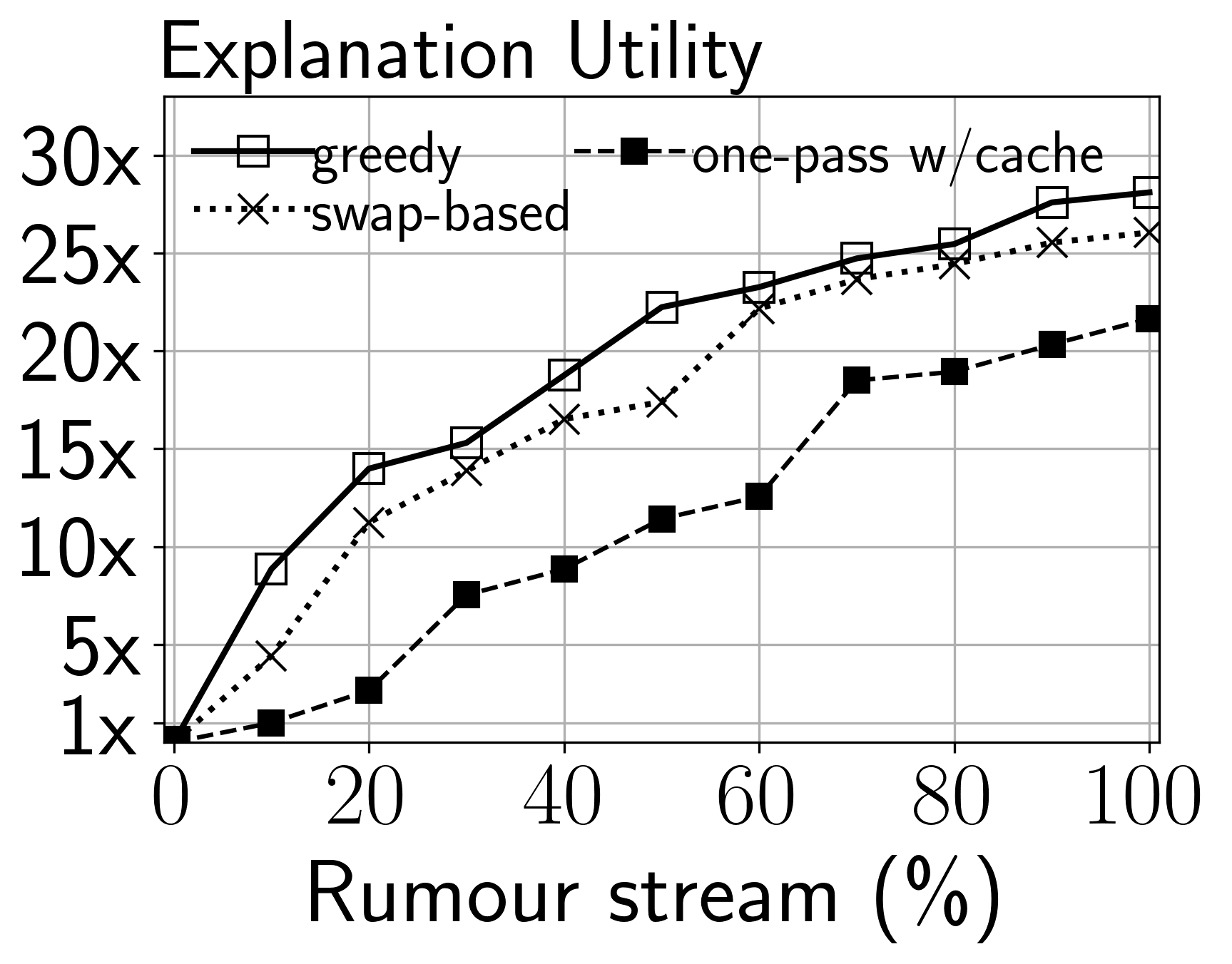}
         \caption{Anomaly dataset}
     \end{subfigure}
\vspace{-.5em}
        \caption{Effects of streaming data on utility}
        \label{fig:explainability_stream}
\vspace{-.5em}
 \end{figure}

\sstitle{Effects of streaming data}
Now, we turn to the change in utility under a stream of rumours. The setting 
is that each arriving rumour is considered as the query, while the explanation 
is derived from all previous rumours. However, to avoid bias, we also select a 
random number of old rumours as 
queries, as in practice, users can freely request an explanation for any 
rumour. 
In other words, we are interested in the cumulative utility of the system 
as a whole rather than the utility for some random query. 
\autoref{fig:explainability_stream} presents the result for Snopes and Anomaly 
datasets (COVID dataset shows a similar trend and is omitted for space 
reasons). We measure the difference of the utility from the beginning of the 
stream to the end. Different 
selection algorithms are used, excluding the \emph{one-pass} algorithm, which 
is not applicable in this scenario as it requires a static query. Once combined 
with caching, it may be used as a list of candidates is maintained for 
each median of rumour clusters (\autoref{sec:caching}).

We observe that the utility increases greatly over time. At the beginning, the 
utility increases slowly, as there is not yet sufficient information for the 
derivation of an explanation. However, the utility improves quickly and 
converges.
For example, in the Snopes dataset, with the \emph{greedy} selection, the 
utility increases $16\times$ with only 50\% of rumours. This 
indicates that the approach can incorporate the information of the multi-modal 
social graph quickly and provide useful explanations. Interestingly, different 
datasets exhibits different scales of utility improvement. For example, the 
\emph{greedy} selection can reach more than $20\times$ improvement with 50\% on 
the Anomaly dataset. This can be attributed to the fact that larger social 
graphs cover more features and structural information, which can be exploited 
to derive explanations.

\subsection{Correctness of Rumour Embedding}
\label{sec:correctness}

Now, we assess whether the embeddings of structurally-similar
subgraphs are indeed close in the embedding space. Our hypothesis is that there is a correlation between the cosine similarity of our graph embeddings and traditional graph similarity measures, including \emph{MCS}, \emph{graphsim}, and \emph{GED}.
To enable the use of \emph{MCS}, for each dataset, we randomly extract 
subgraphs of size 15 from the social graph.
We then select 10K pairs of subgraphs randomly and compute the size of their 
maximum common subgraphs.
In the comparison with \emph{graphsim}, we follow a similar procedure, but 
select pairs of subgraphs of the same size.
As for \emph{GED}, we generate a pair of subgraphs
with edit distance $k$ by first constructing a two-hop ego graph $g$ from a
randomly-selected node in the data graph. We then remove $1\leq k\leq 7$ edges
randomly, so that the subgraph is still connected, to obtain a
subgraph $g'$.
For each pair of subgraphs in every setting, we also measure 
their embedding similarity.

\begin{figure}[!h]
	\centering
	\begin{minipage}{.45\linewidth}
    \centering
    \vspace{-1em}
    \captionsetup{type=table}
    \captionof{table}{Pearson's correlation of embedding similarity and graph 
    similarity}
	\label{tab:correlation}
	\centering
    \scalebox{0.7}{
	\begin{tabular}{l@{\hspace{.2em}}r@{\hspace{.2em}}r@{\hspace{.2em}}r}
		\toprule
		& Snopes & Anomaly & COVID \\ 
		\midrule
		MCS & 0.79   & 0.81   & 0.80         \\
		graphsim & 0.89   & 0.88   & 0.87      \\
        GED & -0.95   & -0.93   & -0.96      \\
		\bottomrule
	\end{tabular}
    }
    \end{minipage}
	\begin{minipage}{0.49\linewidth}
		\centering
		\includegraphics[width=0.6\linewidth]{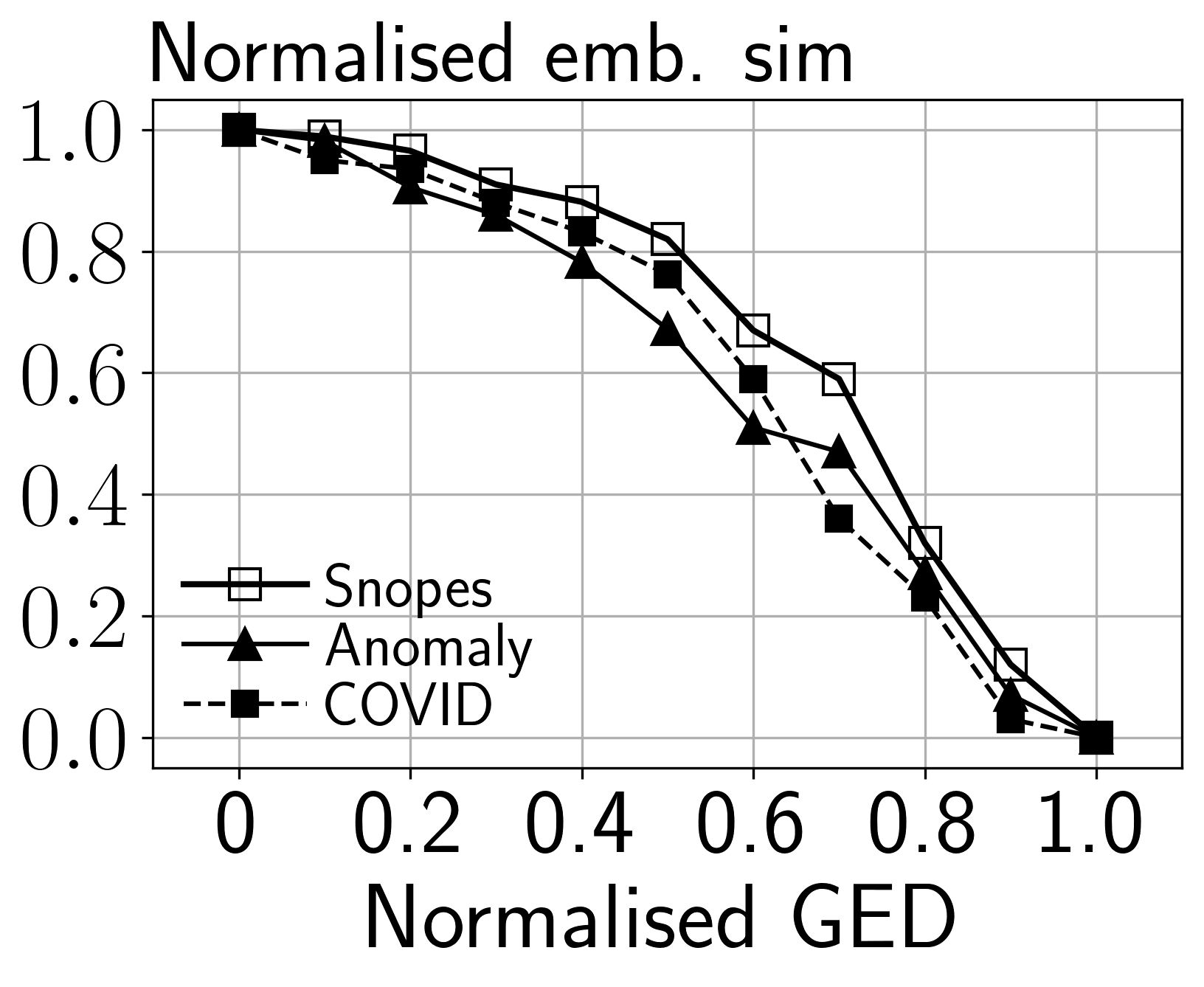}
		\vspace{-1em}
  \caption{Correctness}
  \label{fig:correctness_embedding}
	\end{minipage}
\end{figure}

%

\autoref{tab:correlation} confirms our hypothesis: There is a strong
correlation between our embedding similarity and the graph similarity measures, 
reflected by Pearson's correlation coefficients. \emph{GED} has the strongest 
correlation (a negative correlation, as GED is a distance measure), since, 
similar to our embedding, it incorporates the whole structure of subgraphs, not 
only their common part. \autoref{fig:correctness_embedding} further 
shows the average embedding 
similarity for each GED value. Here, values are normalised to $[0,1]$ 
by dividing by the maximum embedding similarity and the maximum GED value. Over 
all datasets, the GED and the embedding similarity behave consistently. 

\subsection{Efficiency of Rumour Embedding}
\label{sec:efficiency_emb}

\sstitle{Embedding time}
We evaluate the efficiency of our embedding method in 
building the model for the whole social graph (training) as well as indexing a 
new subgraph (testing). To this end, we extract random subgraphs of the 
datasets with the same number of nodes, varying from $10^4$ to $10^6$. Then, we 
measure the training time for each graph and report the average at each graph 
size. 

\begin{figure}[!h]
  \centering
    \begin{minipage}{.47\linewidth}
\centering
 \includegraphics[width=0.7\linewidth]{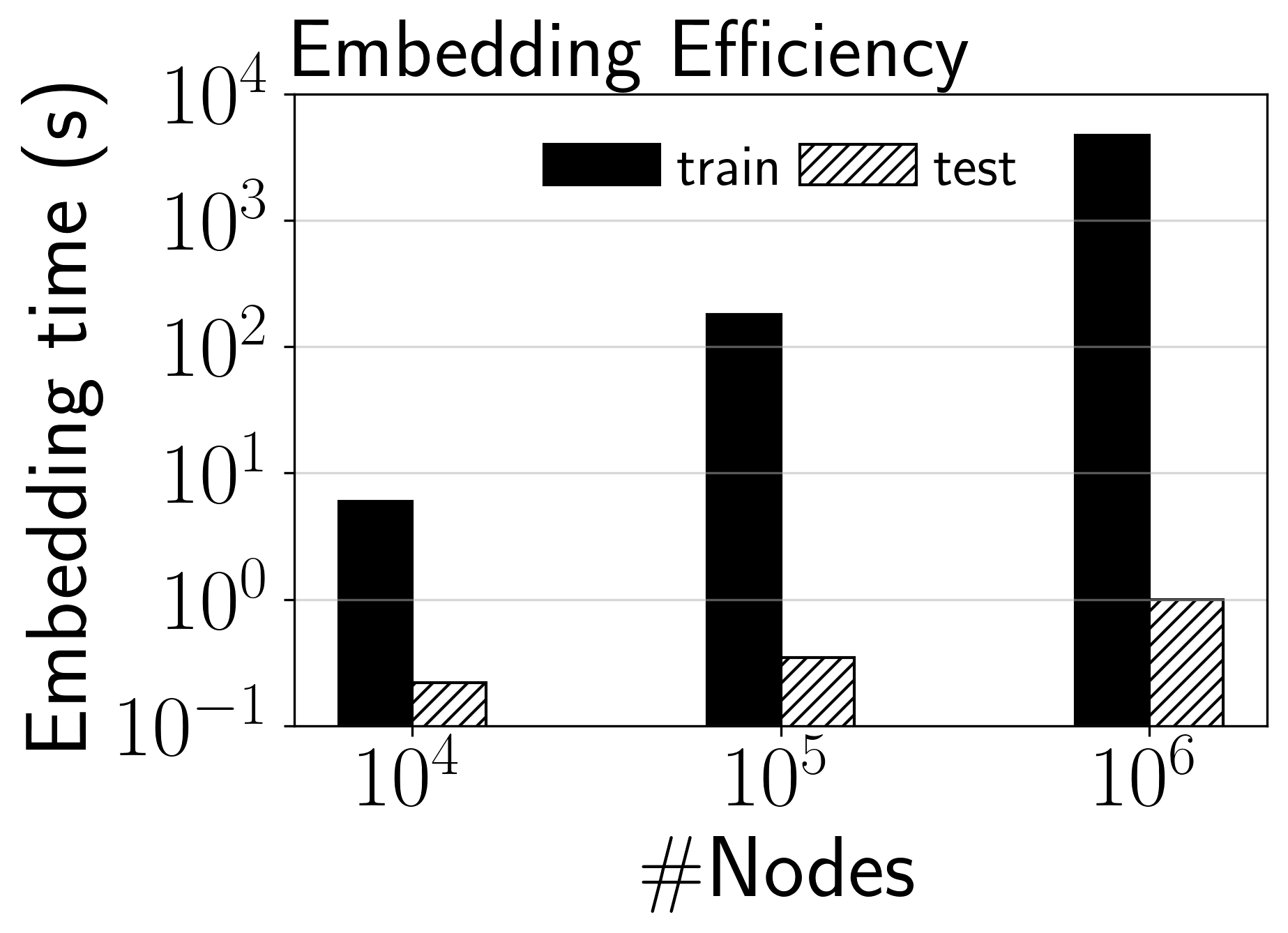}
\vspace{-.5em}
    \caption{Embedding}
  \label{fig:embedding_time}
  \end{minipage}
    \quad
  \begin{minipage}{0.47\linewidth}
    \centering
  	\includegraphics[width=0.7\linewidth]{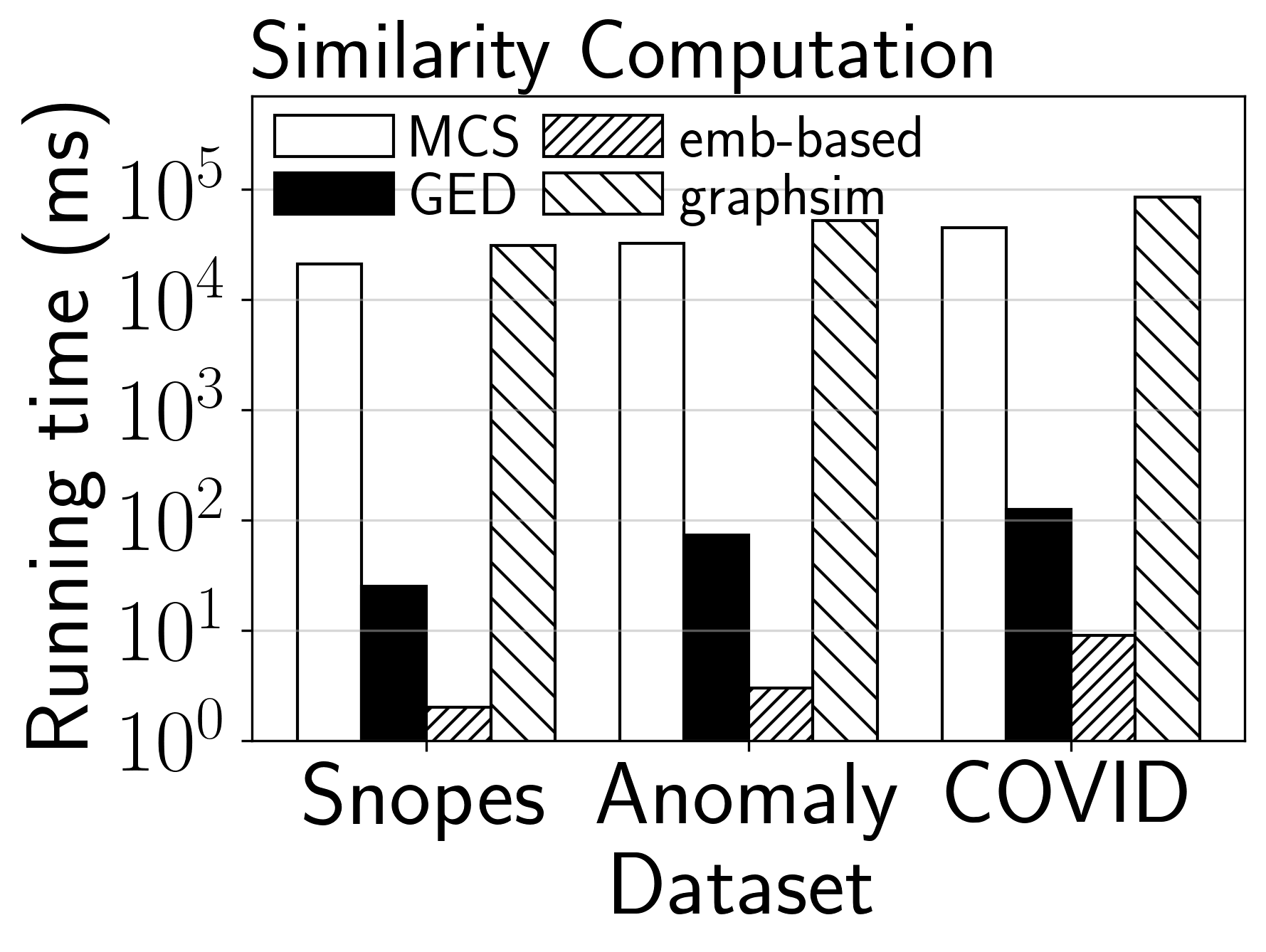}
\vspace{-.5em}
    \caption{Similarity}
  \label{fig:similarity_time}
  \end{minipage}
\vspace{-.5em}
  \end{figure}

As seen in \autoref{fig:embedding_time}, larger graph sizes increase the 
embedding time. Yet, compared to generic graph embedding methods, runtimes are 
competitive~\cite{yang2019aligraphvldb}. In particular, the time to embed a new 
rumour is small ($\leq 1s$ for all cases), as required in streaming settings. 

\sstitle{Similarity computation}
When selecting relevant rumours for explanation, similarity computation is an 
important operation. In this experiment, we evaluate our similarity computation 
method based on embeddings against existing baseline measures, including 
\emph{graphsim}, \emph{MCS}, and \emph{GED}. 
For each dataset, we measure the similarity computation time between all pairs 
of identified rumours and report the average time.

According to \autoref{fig:similarity_time}, our embedding-based method is very 
fast for graph similarity computation, taking only 1ms or less. Moreover, our 
method is 4-5 orders of magnitude faster than \emph{MCS} and 
\emph{graphsim}, which require the 
computation of connected components . 

\sstitle{Indexing performance}
Our embedding method can be seen as a graph indexing technique. Hence, we also 
compare it against graph indices, such as CTIndex or GGSX in terms of time 
required to create an index/vector for a subgraph. For our approach, this is 
the time required to construct an embedding after we already trained the model. 
In this experiment, we set the subgraph size to 32. 


\begin{figure}[!h]
	\centering
\begin{minipage}{0.4\linewidth}
		\centering
    \captionsetup{type=table}
    \captionof{table}{Indexing performance (time$|$space).}
	\label{tab:index}
	\vspace{-1em}
	\scalebox{0.5}{
\begin{tabular}{@{}l c c c@{}}
			\toprule
			& Snopes          & Anomaly           & COVID                 \\ 
			\midrule
			{CTIndex} & 79ms$|$1kB          & 87ms$|$1kB         & 
			1326ms$|$1kB                    \\
			{GGSX}     & 131ms$|$14.01kB          & 528ms$|$232kB          & 
			928ms$|$814kB                   \\
			{Sub2vec} & 1045ms$|$356kB & 1021ms$|$363kB & 1028ms$|$335kB \\
			{Our}    & \textbf{20ms}$|${\bf 1kB} & \textbf{22ms}$|${\bf 1kB} & 
			\textbf{23ms}$|${\bf 1kB}  \\ 
			\bottomrule
		\end{tabular}
	}
	\end{minipage}
	\quad
    \begin{minipage}{.55\linewidth}
    \centering
     \includegraphics[width=1.0\linewidth]{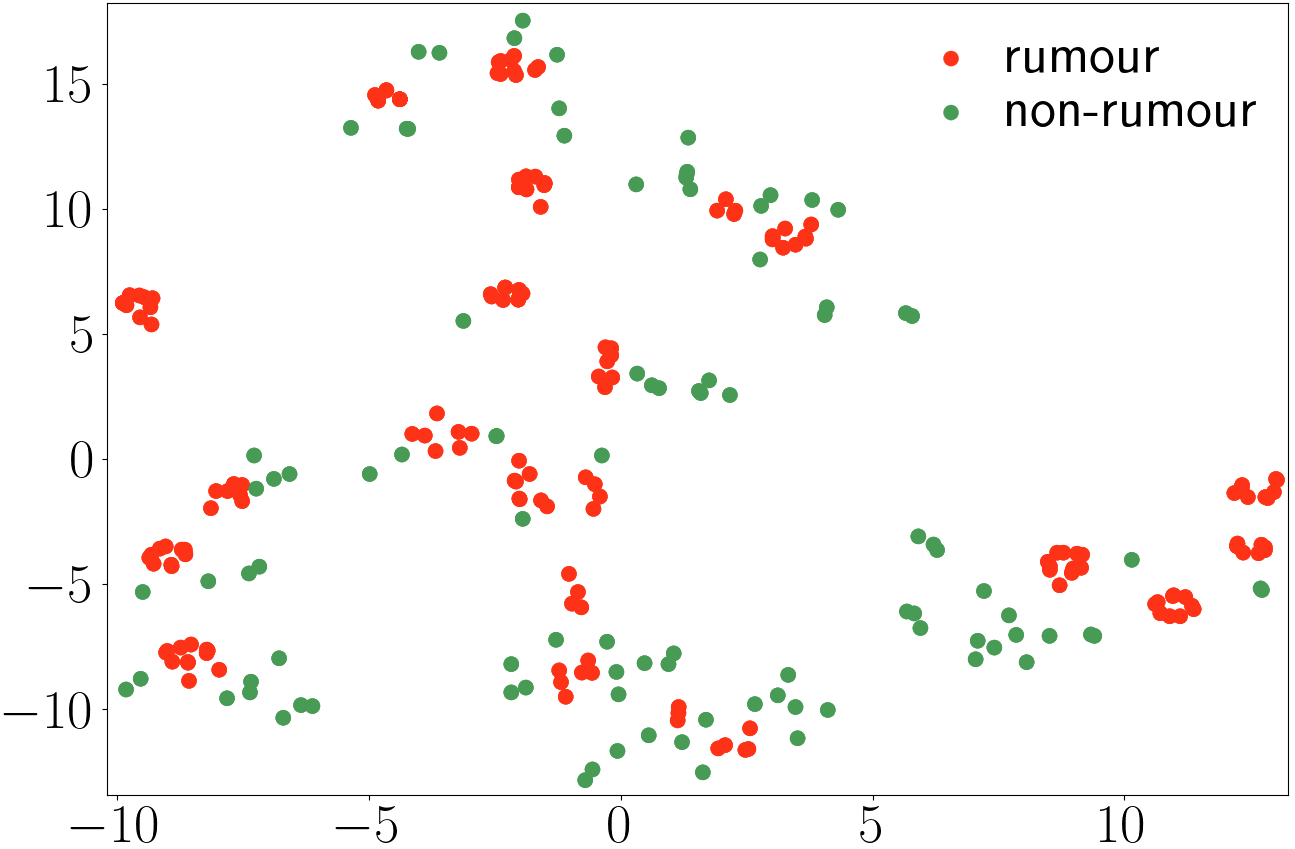}
	\vspace{-1em}
  \caption{t-SNE.}
  \label{fig:qualitative}
	\vspace{-1em}
      \end{minipage}
\end{figure}

\autoref{tab:index} shows the difference in the efficiency of 
structure-based approaches and our embedding-based index. CTIndex requires at 
least 79ms to create an index and on large datasets like \emph{COVID}, it would 
take 1.326s. GGSX performs better than CTIndex for larger datasets like COVID, 
but require 131ms for small datasets like Snopes. Sub2vec, a subgraph embedding 
technique, requires around 1s to generate the embedding for a subgraph. Our 
method is even faster, around one order of magnitude, and constructs an index 
in around 20ms. Note that we evaluate the indexing 
in isolation, without any performance improvements.

Turning to space requirements, we fixed the embedding size for both CTIndex and 
our approach, with a similar index size for a fair comparison. As such, CTIndex 
and our index both require 1kB space. The index size of GGSX, in turn, depends 
on the social graph and ranges from 14.01kB to 814kB. Sub2vec also uses a fixed 
space because the dimensions of the embeddings are often fixed regardless of 
the subgraph size. We conclude that our index has comparable size, but can be 
constructed much quicker.

\subsection{Qualitative Analysis}
\label{sec:quality}

To assess the quality of our rumour embedding model, \autoref{fig:qualitative} 
presents the t-SNE visualisation of the vectors of rumour and non-rumour 
subgraphs in the \emph{Anomaly} dataset (other datasets exhibit similar 
results).  Here, the rumours (red) are clustered and distant from non-rumour 
subgraphs (green). Moreover, there is no single cluster for all rumours, 
implying that we should incorporate not only relevancy but also diversity when 
deriving an explanation.


\subsection{Adaptivity of Explanations}
\label{sec:adaptivity}

Finally, we consider the adaptivity of explanations in cases of concept drifts. 
To simulate such scenarios, we vary the data distribution by randomly 
reordering the appearances of detected rumours in terms of their similarity, 
such that the similarity distribution is \emph{normal} or \emph{skewed}. We 
also vary the input rate (the \emph{high} rate is 5 to 10 times faster than the 
\emph{low} rate). Then, we measure the F1-score as in 
\autoref{sec:effectiveness}. The results are averaged over all datasets and 10 
random runs for each configuration, with an explanation size of seven. 

\begin{table}[!h]
  \centering
  \footnotesize
  \vspace{-0.5em}
 \caption{Explanation accuracy in adverse conditions}
  \label{tab:adaptive}
  \vspace{-.5em}
  \begin{tabular}{clccc}
    \toprule
     Data & Input & Static & Dynamic & Dynamic embedding \\
     dist. & rate & embedding & embedding & w/ drift detection    \\
    \midrule
    \multirow{2}{*}{normal} & low & \colorbox{orange}{+0.00\%} & \colorbox{lime}{+45.32\%} & \colorbox{green}{+67.95\%}  \\
     & high & \colorbox{orange}{+0.00\%} & \colorbox{lime}{+45.04\%} & \colorbox{green}{+67.95\%}  \\
    \midrule
    \multirow{2}{*}{skewed} & low & \colorbox{red}{-23.25\%} & \colorbox{yellow}{+33.98\%} & \colorbox{lime}{+61.79\%}  \\
     & high & \colorbox{red}{-23.25\%} & \colorbox{yellow}{+22.95\%} & \colorbox{lime}{+61.79\%}  \\
    \bottomrule
  \end{tabular}
   \vspace{-0.25em}
\end{table}

\autoref{tab:adaptive} presents the results for three versions of our 
framework: \emph{static embedding}, a baseline where the model is not updated; 
\emph{dynamic embedding}, where the model is updated as explained in 
\autoref{sec:drift} after a fixed amount of time; and \emph{dynamic embedding 
w/ drift detection}, where the model is updated when a concept drift is 
detected as presented in \autoref{sec:drift}. We report the difference in the 
F1-score from the baseline. Dynamic embedding turns out to improve 
the performance in adverse conditions, while drift detection further improves 
the robustness to different input rates and data distributions.

\section{Conclusions}
\label{sec:conclusion}

With the detection of rumours on social media becoming an increasingly 
important task, in this paper, we presented an approach to help users 
understand why certain entities are classified as a rumour. To this end, we 
followed the paradigm of example-based explanations of rumours for a 
graph-based model. That is, given a rumour graph, our approach extracts the $k$ 
most similar subgraphs that represent past rumours, while ensuring a certain 
level of diversity in the result. To achieve efficient construction of 
explanations, we proposed a novel technique for graph representation learning 
that incorporates both, features and modalities of nodes and edges. Further 
optimizations include indexing and caching schemes, as well as means to adapt 
to concept drift. 
Our experiments with diverse datasets highlight the efficiency 
and effectiveness of our approach. Especially, our approach yields explanations 
of much higher accumulated utility, up to a factor of $16$, than baseline 
techniques.
This implies that users would receive more information when investigating rumours with our explanation framework.

One limitation of our work is that explanations are based on real samples, so that some users might find it difficult to generalise from them. One possible improvement is include synthetic samples as additional explanation for users to compare, which however might overwhelm user cognitive load.
In future work, we plan to use our explanations for downstream 
applications, such as group attack detection. Also, we intend to investigate the consensus computing of user feedback on explanation utility~\cite{nimmy2022explainability} and the integration of pseudo feedback mechanisms~\cite{gan2018extracting,zhao2021scalable}.






%

\end{document}